\documentclass[runningheads]{llncs}
\usepackage[T1]{fontenc}
\usepackage{graphicx}
\usepackage[margin=1.3in]{geometry}
\usepackage{amsmath,amsfonts,amssymb}
\usepackage{booktabs}
\usepackage{pifont}
\usepackage[ruled,linesnumbered,vlined]{algorithm2e}
\usepackage{float}
\usepackage{mathrsfs}
\usepackage{multirow,multicol}
\usepackage{subfigure}
\usepackage{enumitem}
\usepackage[all]{xy}
\usepackage{xcolor}
\usepackage{bbm}
\usepackage{makecell}
\usepackage{pdfpages}
\usepackage{hyperref}
\hypersetup{colorlinks=true, citecolor=blue, linkcolor=red}

\def\no{\ding{55}}

\def\p{\mathbb{P}}
\def\R{\mathcal{R}}

\def\LA{\mathcal{L}(A)}
\def\erm{ {\rm e} }

\def\CW{\operatorname{CW}}
\def\CL{\operatorname{CL}}
\def\CL{\operatorname{CL}}

\def\borda{\operatorname{\mathsf{Borda}}}

\newcommand{\calo}{\mathcal{O}}

\newenvironment{pfsketch}{\noindent{\em Proof Sketch.}\hspace*{0.5em}}{\hfill{\qed}\vspace{0.8ex}}
\newcommand{\parahead}[1]{\vspace{0.8ex} \noindent {\bf #1.}~}

\SetAlgorithmName{Mechanism}

\begin{document}
\title{Trading Off Voting Axioms for Privacy}

%\titlerunning{Abbreviated paper title}
% If the paper title is too long for the running head, you can set
% an abbreviated paper title here
%
    \author{Zhechen Li\inst{1} \and
        Ao Liu\inst{2} \and
        Lirong Xia\inst{3} \and
        Yongzhi Cao\inst{1} \and
        Hanpin Wang\inst{1,4}
        }

    \authorrunning{Z. Li et al.}
    % First names are abbreviated in the running head.
    % If there are more than two authors, 'et al.' is used.
    %
    \institute{1Key Laboratory of High Confidence Software Technologies (MOE), School of Computer Science, Peking University, China \\
    \email{\{lizhechen, caoyz, whpxhy\}@pku.edu.cn} \and
    Google LLC \\
    \email{aoliu.cs@gmail.com} \and
    Department of Computer Science, Rensselaer Polytechnic Institute \\ 
    \email{xialirong@gmail.com} \and
    School of Computer Science and Cyber Engineering, Guangzhou University, China
    }

\maketitle

\begin{abstract}
    In this paper, we investigate tradeoffs among differential privacy (DP) and several important voting axioms: Pareto efficiency, SD-efficiency, PC-efficiency, Condorcet consistency, and Condorcet loser criterion.  We provide upper and lower bounds on the two-way tradeoffs between DP and each axiom. We also provide upper and lower bounds on three-way tradeoffs among DP and every pairwise combination of all the axioms, showing that, while the axioms are compatible without DP, their upper bounds cannot be achieved simultaneously under DP. Our results illustrate the effect of DP on the satisfaction and compatibility of voting axioms.
\end{abstract}

\section{Introduction}
Voting is a popular method for collective decision making. In a typical voting process, voters express their preferences over the alternatives, and a winner is determined according to a voting rule. In modern social choice theory, voting rules are evaluated and compared by the axiomatic approach~\cite{Plott76:Axiomatic}, w.r.t.~their satisfaction of various normative properties, known as voting axioms. For instance, Pareto efficiency requires that any Pareto-dominated alternative (i.e., an alternative considered worse than another alternative by all voters) should never be selected as the winner.

% These axioms can be roughly divided into four parts~\cite{Brandt2017:Rolling}, namely efficiency, consistency, strategyproofness, and participation. All of them play a crucial role in determining the quality and desirability of a voting rule.\lirong{not sure if we should mention the four parts: we only solved two. Plus I am not sure if this categorization is accurate. We can probably remove the last two sentences.}

\emph{Differential privacy} (DP) is a \emph{de facto} notion of privacy and is widely used in the field of machine learning~\cite{abadi2016deep,vasa2023deep,sarwate2013signal}, data mining~\cite{friedman2010data,zhang2011distributed}, and recommendation systems~\cite{berlioz2015applying,li2020federated}, just to name a few.  In recent years, privacy has also become a major concern in voting schemes, prompting extensive research on differential privacy in the context of voting and rank aggregation~\cite{shang2014application,hay2017differentially,yan2020private,ao2020private}. Those works focus on the privacy-utility tradeoff in voting rules (or rank aggregations), where the utility usually is measured by accuracy or mean square error. Even though some previous work~\cite{DBLP:conf/ijcai/Lee15,li2022differentially} has explored the tradeoff between DP and certain voting axioms through several mechanisms, the overall tradeoff between DP and voting axioms remains largely unexplored. Besides, the relationship between voting axioms may be affected by DP. For example, Condorcet criterion is compatible with Pareto efficiency in social choice theory, where Condorcet criterion requires that the Condorcet winner (the alternative defeats all components in pairwise comparison) must win the election. However, when DP is required, Condorcet criterion and Pareto efficiency are no longer compatible. At a high level, this is because DP requires the voting rules to be random, i.e., all alternatives need to have a non-zero probability to win. Thus, under DP, (better) Condorcet criterion requires the Condorcet winner to win with a larger probability, while (better) Pareto efficiency focus on the pairwise difference in winning probability (Pareto-dominate alternative need to have a higher probability to win compared to the Pareto-dominated alternative). Till now, the incompatibility between axioms brought by DP has not been studied. Thus, the following question still remains largely open.

\begin{center}
    \em What is the tradeoff among DP and voting axioms?
\end{center}

%This paper focuses on the axioms about efficiency and Condorcet consistency \ao{not sure if it's the correct word.} 
Since DP requires randomness in the voting rule, the (strict) axioms are usually not compatible with DP \cite{li2022differentially}. Thus, we propose approximate voting axioms and study the tradeoff among DP and those approximate axioms. %We apply the same approximate Condorcet criterion as \cite{li2022differentially}. In addition, we propose approximate Pareto efficiency, %(Definition \ref{def: beta-pareto}), approximate SD-efficiency, approximate PC-efficiency, and approximate Condorcet loser criterion. 
% \ao{Not sure if listing this as a conceptual contribution will be better?} \lirong{agreed. we should just say that only approximate Condorcet was defined in previous work (if true), and then discuss the new ones in the beginning of the next subsection}

\subsection{Our contributions}

Our conceptual contribution involves proposing approximate versions of Pareto efficiency, SD-efficiency, PC-efficiency, and Condorcet loser criterion (Definitions \ref{def: beta-pareto}-\ref{def: eta-Condorcet-loser})%. These approximate axioms are applied to measure the tradeoffs discussed in the paper.

\begin{figure}
    \centering
    \includegraphics[width=.9\linewidth]{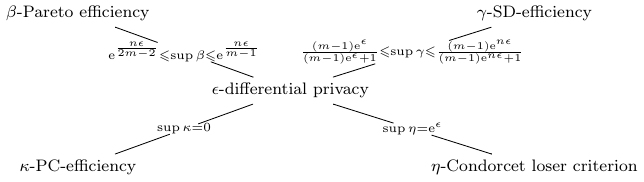}
    \caption{Tradeoff between $\epsilon$-DP and voting axioms.}
    \label{fig: diagram-2tradeoff}
\end{figure}

Our theoretical contributions are two-fold. Firstly,  we explore the 2-way tradeoff between DP and a single (approximate) axiom. We provide tradeoff theorems (with both upper and lower bounds) between DP and the approximate Pareto, SD, and PC efficiencies (Propositions~\ref{prop: tradeoff-DP-Pareto}-\ref{prop: tradeoff-DP-PCeff}). For Condorcet criterion and Condorcet loser criterion, we provide a tight bound for their tradeoffs with DP (Propositions~\ref{prop: CW-DP-lower}-\ref{prop: CL-DP-lower}). We summarize this part of contributions in Fig. \ref{fig: diagram-2tradeoff}, where the expressions displayed on the lines illustrate the bounds of the approximate axioms under $\epsilon$-DP.

% \lirong{I don't see a solid line in the figure. We probably don't need to mention them in the caption (but need to mention in Our Contributions).}

\begin{figure}
    \centering
    \includegraphics[width=.9\linewidth]{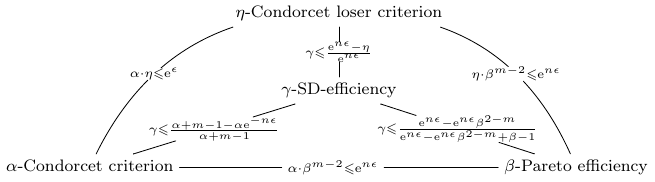}
    \caption{Relations between voting axioms under $\epsilon$-DP.}
    \label{fig: diagram-3tradeoff}
    % \vspace{-1.5em}
\end{figure}

Secondly, we explore the 3-way tradeoff between DP and two axioms (also called tradeoff axioms under DP in this paper). Since even the approximate version of PC-efficiency cannot be achieved under DP, we consider Condorcet criterion, Condorcet loser criterion, Pareto efficiency, and SD-efficiency in 3-way tradeoffs. We capture the lower bounds of 3-way tradeoffs through several mechanisms, summarized by Table \ref{tab: 3way-tradeoff-lower} in Section \ref{sec: 3-way-tradeoff}. Then we investigate all of their pairwise tradeoff under DP by proving their upper bounds (Theorems \ref{thm: Condorcet-winner-loser}-\ref{thm: SD-PE}). We show that there exists a tradeoff between each pair of these axioms, though they are compatible without DP. This part of results is summarized in Fig. \ref{fig: diagram-3tradeoff}, where the expressions displayed on the lines illustrate the upper bounds of 3-way tradeoffs between the axioms.

% \lirong{again, in Fig 2 caption, no need to mention ``solid line''.}

Algorithmically, we propose a group of (randomized) voting rules able to make smooth 2-way tradeoffs between DP and the axioms mentioned above (Mechanisms \ref{algo: BordaEXP}-\ref{algo: CL-RR}). For the 3-way tradeoffs, we propose a mechanism able to flexibly tradeoff between Condorcet criterion and Condorcet loser criterion (Mechanism \ref{algo: CWRR-CLRR}). Finally, we experimentally verified the tradeoffs in the proposed mechanisms.

%Experimentally, \ao{introduce experiments here.} \zc{Not sure if the experiments should be emphasized here.}

\subsection{Related work and discussions}

To the best of our knowledge, the application of DP to rank aggregations (a generalized problem of voting) was first introduced in~\cite{shang2014application}, which derived upper bounds on error rate under DP. 
In a similar vein, Lee~\cite{DBLP:conf/ijcai/Lee15} proposed a tournament voting rule that achieves both DP and robustness to strategic manipulations.
Hay et al.~\cite{hay2017differentially} used the Laplace mechanism and exponential mechanism to enhance the privacy of Quicksort and Kemeny-Young methods.
Kohli and Laskowski~\cite{kohli2018epsilon} investigated DP, strategyproofness, and anonymity for voting on single-peaked preferences. 
Torra~\cite{torra2019random} analyzed the privacy-preserving level of the random dictatorship, a well-known randomized voting rule, using DP. Their study examined the conditions under which random dictatorship satisfies DP and proposed improvements to achieve it in general cases.
Wang et al.~\cite{wang2019aggregating} analyzed the privacy of positional voting and proposed a new noise-adding mechanism that outperforms the naive Laplace mechanism in terms of accuracy. 
Yan et al.~\cite{yan2020private} addressed the tradeoff between accuracy and privacy in rank aggregation by achieving local DP through Laplace mechanism and randomized response. They explored mechanisms that strike a balance between privacy and utility in the context of rank aggregation.

Most of the aforementioned works did not consider the tradeoff between privacy and voting axioms. The privacy bounds in those works are usually not tight. Liu et al.~\cite{ao2020private} introduced distributional DP~\cite{bassily2013coupled} to voting and studied the privacy level of several commonly used voting rules, but they did not provide any approach to improve privacy. Li et al.~\cite{li2022differentially} introduced a novel family of voting rules that satisfy DP along with various voting axioms. They also explored the tradeoff between DP and several axioms, including Pareto efficiency, Condorcet criterion, and others. However, Li et al.~\cite{li2022differentially} only briefly discussed the general tradeoff between DP and certain axioms, and their discussion was limited to Condorcet criterion and Pareto axioms. A more comprehensive and in-depth analysis is still required in this regard. Beyond social choice, DP has also been considered in other topics of economics, such as mechanism design~\cite{pai2013privacy,xiao2013privacy}, matching~\cite{hsu2016private}, and resource allocation~\cite{hsu2016private,kannan2018private}.

In the context of social choice theory, there is a large literature on the analysis of randomized voting~\cite{Brandt2017:Rolling}. Most of them focused on assessing the satisfaction of standard axiomatic properties, such as the complexity of manipulation~\cite{walsh2012lot}, strategyproofness~\cite{aziz2014incompatibility,aziz2015universal}, Pareto efficiency~\cite{brandl2015incentives,gross2017vote}, participation~\cite{brandl2019welfare} and monotonicity~\cite{DBLP:conf/ijcai/Brandl0S18}. In sortition, the fairness properties have also been investigated~\cite{benade2019no,flanigan2020neutralizing,flanigan2021fair}. For approximate axiomatic properties, Procaccia~\cite{procaccia2010can} discussed how much a strategyproof randomized rule could approximate a deterministic rule. Birrell and Pass~\cite{birrell2011approximately} explored the approximate strategyproofness for randomized voting rules. All those approximate axioms are not natural in the context of DP because approximate axioms adopt the difference in utility functions but DP focuses on ratios.

\section{Preliminaries}
\label{sec: prelim}

Let $A=\{a_1,a_2,\ldots,a_m\}$ denote a set of $m\geqslant 2$ alternatives.
For any $n\in\mathbb N$, let $N=\{1,2,\ldots,n\}$ be a set of voters.
For each $j\in N$, the vote of voter $j$ is a linear order $\succ_j\in\LA$, where $\LA$ denotes the set of all linear orders over $A$, i.e., all transitive, anti-reflexive, anti-symmetric, and complete binary relations.
$P=\{\succ_1,\succ_2,\ldots,\succ_n\}$ denotes the \emph{(preference) profile}, which is the collection of $n$ votes. For any $j\in N$, let $P_{-j} = \{\succ_1,\ldots,\succ_{j-1},\succ_{j+1},\ldots\succ_n\}$ denote the profile of removing the $j$-th vote from $P$.

Under the settings above, a {\em (randomized) voting rule} can be defined as a mapping $f\colon \LA^n\to \R(A)$, where $\R(A)$ denotes the set of all random variables on $A$ (usually called {\em lotteries} in social choice theory). Given a voting rule $f$ and a profile $P$, the winning probability of alternative $a\in A$ is denoted by $\p[f(P)=a]$. A voting rule $f$ is {\em neutral} if for any profile $P$ and any permutation $\sigma$ on $A$, $\sigma\cdot f(P)=f(\sigma\cdot P)$.

\parahead{Differential privacy (DP\footnote{We use DP to represent either \emph{differential privacy} or \emph{differentially private} throughout this paper.}, ~\cite{Dwork06D})} At a high level, DP requires a function to have similar output (distribution) when inputting {\em neighboring databases}. Here, we say two databases are \emph{neighboring} if one can be gotten by replacing no more than one entry from the other database.

\begin{definition}[\bf\boldmath Differential Privacy~\cite{Dwork06D}]\label{P}
    \label{def: dp}
    A mechanism $f:\mathcal{D} \to \calo$ satisfies $\epsilon$-differential privacy ($\epsilon$-DP for short) if for all $O\subseteq \calo$ and each pair of neighboring databases $P,P'\in \mathcal{D}$,
    \begin{align*}
        \p[f(P)\in O] \leqslant \erm^{\epsilon}\cdot \p[f(P')\in O].
    \end{align*}
    % where the probability comes from the randomness in function $f$.
\end{definition}
The probability of the above inequality is taken over the randomness of the mechanism. The smaller $\epsilon$ is, the better privacy guarantee can be offered. In the context of social choice, the mechanism $f$ in Definition \ref{def: dp} is a voting rule and its domain
\begin{align*}
    \mathcal{D}=\LA^*=\LA\cup \LA^2\cup \cdots.
\end{align*}
Further, the pair of neighboring databases $P,P'$ are two profiles differing on no more than one vote, i.e., there exists a voter $j\in N$ such that $P_{-j}=P_{-j}'$.

\parahead{Axioms of voting} The axioms considered in the paper can be roughly divided into two parts: axioms that only depend on preferences over alternatives and axioms that depend on preferences over lotteries in $\R(A)$. We adopt the notions in \cite{Brandt2017:Rolling} here.
%  \ao{Get confused here, what means ``preferences over lotteries''? Is ``lottery'' the output of the voting rule?}

For clarity, we define the following notions before presenting the axioms. Given a profile $P$ and two distinct alternatives $a,b\in A$, let $w_P[a,b]$ denote the {\em majority margin} of $a$ over $b$, which equals to the number of voter that consider $a\succ b$ minus the number of voter that consider $b\succ a$, i.e.,
\begin{align*}
    w_P[a,b] = |\{j\in N: a\succ_j b\}| - |\{j\in N: b\succ_j a\}|.
\end{align*}
Here, we slightly abuse the notations and let $a\succ_j b$ represent that the $j$-th voter prefers $a$ to $b$. Again, $\succ_j$ represents the $j$-th vote (a linear order) when appearing alone. The {\em Condorcet winner} of $P$ (denoted as $\CW(P)$) is an alternative $a\in A$ such that $w_P[a,b]>0$ for all $b\neq a$. Similarly, the {\em Condorcet loser} of $P$ (denoted as $\CL(P)$) is the alternative $a\in A$ which satisfies $w_P[a,b] < 0$ for all $b\neq a$. Based on these notions, the first part of the axioms are listed below.

\begin{itemize}
    \item {\em Condorcet criterion}: A voting rule $f$ satisfies Condorcet criterion if $\p[f(P)=\CW (P)]=1$ holds for all profile $P$ that $\CW(P)$ exists;
    \item {\em Condorcet loser criterion}: A voting rule $f$ satisfies Condorcet loser criterion if $\p[f(P)=\CL (P)]=0$ holds for all profile $P$ that $\CL(P)$ exists;
    \item {\em Pareto efficiency}: A voting rule $f$ satisfies Pareto efficiency if $\p[ f(P)= b]=0$ for all profile $P\in\LA^n$, where $b\in A$ is Pareto dominated by some $a\in A$, i.e., $a\succ_j b$ for all $j\in N$.
\end{itemize}

The second part of axioms (efficiency notions except Pareto efficiency) considers the relationship between lotteries. For clarity, we define the following relationships before presenting the axioms.\\

{\em Stochastic Dominance (SD)}: Given the $j$-th vote $\succ_j\,\in\LA$ and two lotteries $\xi,\zeta\in \R(A)$, $\xi$ is more desirable under SD for the $j$-th voter (denoted by $\xi \succeq_j^{SD} \zeta$) if and only if for every $x\in A$, the probability of $\xi$ selecting an alternative better than $x$ is no less than the probability for $\zeta$ to select such an alternative, i.e.,
\begin{align}
    \label{equ: SD-def}
    \sum\limits_{x\succ y} \p[\xi = x] \geqslant \sum\limits_{x\succ y} \p[\zeta = x], \quad\text{for all }y\in A.
\end{align}
We say $\xi$ is strictly more desirable than $\zeta$ by means of SD for the $j$-th voter  (denoted by $\xi\succ_j^{SD} \zeta$) if and only if $\xi\succeq_j^{SD}\zeta$ holds and $\zeta\succ_j^{SD}\xi$ does not hold.

    % and satisfy
    % \begin{align*}
    %     \sum\limits_{y\succ x} \p[\xi = x] > \sum\limits_{y\succ x} \p[\zeta = x], \quad\text{for some }x\in A.
    % \end{align*}
    % \ao{Several questions here: 1. Please add a sentence (in words) to explain the axiom. 2. $\succ\,\in \LA$ looks confusing, I suggest changing it to $\succ\,\in \LA$ throughout this paper. 3. The meaning of $\succeq^{SD}$ need to be defined before using. }
    {\em Pairwise Comparison (PC)}: Given the $j$-th vote $\succ_j\,\in\LA$ and two lotteries $\xi,\zeta\in \R(A)$, $\xi$ is more desirable under PC for the $j$-th voter (denoted by $\xi \succeq_j^{PC} \zeta$) if and only if the probability that $\xi$ yields a better alternative than $\zeta$ is no less than the other way round, i.e.,
\begin{align}
    \label{equ: PC-def}
    \sum\limits_{x\succ y} \p[\xi = x] \cdot \p[\zeta = y] \geqslant \sum\limits_{x\succ y} \p[\zeta = x] \cdot \p[\xi = y].
\end{align}
Similarly, $\xi$ is strictly more desirable than $\zeta$ by means of PC for the $j$-th voter (denoted by $\xi\succ_j^{PC} \zeta$) if and only if $\xi\succeq_j^{PC}\zeta$ holds and $\zeta\succ_j^{PC}\xi$ does not hold.\\

Based on SD and PC, the second part of axioms (SD-efficiency and PC-efficiency) are defined as follows.

\begin{itemize}
    \item {\em SD-Efficiency: }A voting rule $f$ satisfies {\em SD-efficiency} if for all profile $P\in \LA^n$, there does not exist $\xi\in \R(A)$ satisfying both of the following two conditions.

          \qquad 1. For all $j\in N$, $\xi\succeq_j^{SD} f(P).$ \qquad\quad 2. There exists some $j\in N$ that $\xi\succ_j^{SD} f(P)$.\\

    \item {\em PC-Efficiency: } A voting rule $f$ satisfies {\em PC-efficiency} if for all profile $P\in \LA^n$, there does not exist $\xi\in \R(A)$ satisfying both of the following two conditions.

          \qquad 1. For all $j\in N$, $\xi\succeq_j^{PC} f(P).$ \qquad\quad 2. There exists some $j\in N$ that $\xi\succ_j^{PC} f(P)$.\\
\end{itemize}

The relationship among the notions of efficiency mentioned in our paper can be visualized as the following diagram, where $a\to b$ indicates that $a$ implies $b$.
\begin{align*}
    \text{(Strongest) PC-efficiency} \to \text{SD-efficiency} \to \text{Pareto~efficiency (Weakest)}
\end{align*}

% \ao{Did we used $\mathcal{E}$-Efficiency anywhere? If not, we can just remove the following definition and move the rest part to where it is first used.}

% Given a lottery extension $\mathcal{E}$ \ao{what is lottery extension? I think we need to make either examples or explanations here.}, the strict part \ao{what means strict part?} of $\succeq^{\mathcal{E}}$ is defined as $\xi \succ^{\mathcal{E}} \zeta$ if and only if $\xi\succeq^{\mathcal{E}} \zeta$ and not $\zeta\succeq^{\mathcal{E}} \xi$. \ao{Why introduce $\succeq^{\mathcal{E}}$? I think $\mathcal{E}$-Efficiency is defined upon it, am I correct?} Under the settings above, efficiency can be defined according to the lottery extensions above.

% \begin{definition}[\bf\boldmath $\mathcal{E}$-Efficiency]
%     Given a lottery extension $\mathcal{E}$, a voting rule $f$ satisfies {\em $\mathcal{E}$-efficiency} if for all profile $P\in \LA^n$, there does not exist $\xi\in \R(A)$ that $\mathcal{E}$-dominates $f(P)$, i.e., $\xi\succeq_j^{\mathcal{E}} f(P)$ for all $j\in N$ and $\xi\succ_j^{\mathcal{E}} f(P)$ for some $j\in N$.
% \end{definition}

\section{DP-Axioms Tradeoff}
This section investigates the tradeoff between privacy and voting axioms (2-way tradeoff). The axioms considered here can be divided into two parts, efficiency (Pareto efficiency, SD-efficiency, and PC-efficiency), and Condorcet consistency (Condorcet criterion and Condorcet loser criterion). Table~\ref{tab: privacy-efficiency} summarizes the theoretical results of this section. As all five axioms are not compatible with DP, we propose their approximate version (if has not yet been proposed in literature). With the approximate axioms, we establish both upper and lower bounds about their tradeoffs with DP, i.e., the upper and lower bound of approximate axioms with a given privacy budget $\epsilon$. All missing proofs of this section can be found in Appendix A.

\begin{table}
    \caption{The tradeoff between DP and voting axioms, where \no~indicates that the standard efficiency is incompatible with DP. The expressions in the table represent the level of satisfaction to approximate notions of efficiency under $\epsilon$-DP (approximate PC-efficiency is not achievable under DP, so there is no lower bound here).}
    \label{tab: privacy-efficiency}
    \centering
    \setlength\tabcolsep{5pt}
    \begin{tabular}{ccccc}
        \toprule
        Voting axiom              & Compatibility & Upper bound                                             & Lower bound                                           & Reference                                                              \\

        \midrule
        Pareto efficiency         & \no           & $\erm^{\frac{n\epsilon}{m-1}}$                          & $\erm^{\frac{n\epsilon}{2m-2}}$                       & Propositions \ref{prop: tradeoff-DP-Pareto}-\ref{prop: bordaExp-pareto} \\
        SD-efficiency             & \no           & $\frac{(m-1)\erm^{n\epsilon}}{(m-1)\erm^{n\epsilon}+1}$ & $\frac{(m-1)\erm^{\epsilon}}{(m-1)\erm^{\epsilon}+1}$ & Propositions \ref{prop: tradeoff-DP-SD-eff}-\ref{prop: anti-exp-SDeff}  \\
        PC-efficiency             & \no           & $0$                                                     & ---                                                   & Proposition \ref{prop: tradeoff-DP-PCeff}                              \\
        Condorcet criterion       & \no           & $\erm^{\epsilon}$~\cite{li2022differentially}           & $\erm^{\epsilon}$                                     & Proposition \ref{prop: CW-DP-lower}                                    \\
        Condorcet loser criterion & \no           & $\erm^{\epsilon}$                                       & $\erm^{\epsilon}$                                     & Propositions \ref{prop: CL-DP-upper}-\ref{prop: CL-DP-lower}            \\
        \bottomrule
    \end{tabular}
\end{table}

\subsection{DP-Efficiency tradeoff}
First of all, we discuss the tradeoff between DP and Pareto efficiency. Li et al.~\cite{li2022differentially} introduced the concept of probabilistic Pareto efficiency to address the inherent incompatibility between DP and Pareto efficiency. Probabilistic Pareto efficiency stipulates that each Pareto dominating alternative must have a higher probability of winning compared to the dominated alternative (standard Pareto efficiency requires ``always winning'' instead of ``a higher probability of winning''). We further extend this notion by introducing a parameter $\beta$ to quantify the level of Pareto efficiency.
\begin{definition}[\bf\boldmath $\beta$-Pareto Efficiency]
    \label{def: beta-pareto}
    Given $\beta>0$, a voting rule $f\colon \LA^n\to \R(A)$ satisfies $\beta$-Pareto efficiency, if for each pair of alternatives $a,b\in A$ that $a\succ_j b$ for all $j\in N$, it holds that
    \begin{align*}
        \p[f(P)=a]\geqslant \beta\cdot \p[f(P)=b].
    \end{align*}
\end{definition}

In words, a voting rule satisfies $\beta$-Pareto efficiency if the winning probability of each Pareto dominated alternative never exceeds $1/\beta$ of the winning probability of its dominant alternative. Therefore, in Definition \ref{def: beta-pareto}, a larger $\beta$ is more desirable and represents a higher level of Pareto efficiency. It's easy to check that $1$-Pareto efficiency is equivalent to the probabilistic Pareto efficiency in~\cite{li2022differentially} and $\infty$-Pareto efficiency is equivalent to standard Pareto efficiency.

Next, we tradeoff Pareto efficiency with DP under the notion of $\beta$-Pareto efficiency. The following proposition provides an upper bound of $\beta$-Pareto efficiency under the constraint of $\epsilon$-DP.
\begin{proposition}[\bf\boldmath $\beta$-Pareto Efficiency, Upper Bound]
    \label{prop: tradeoff-DP-Pareto}
    There is no neutral rule $f\colon \LA^n\to \R(A)$ that satisfies $\epsilon$-DP and $\beta$-Pareto efficiency with $\beta> \erm^{\frac{n\epsilon}{m-1}}$.
\end{proposition}
\begin{pfsketch}
    Let $f$ be a neutral rule satisfying $\epsilon$-DP and $\beta$-Pareto efficiency. By $\epsilon$-DP and neutrality, we can prove that
    \begin{align}
        \label{equ: n-eps-pareto-bound}
        \p[f(P)=a] & \leqslant \erm^{n\epsilon}\cdot \p[f(P)=b],\quad \text{for all }P\in\LA^n \text{ and } a,b\in A.
    \end{align}
    Then, by considering the profile $P$  where all voters' preferences are the same, i.e., $a_1\succ_j a_2\succ_j\cdots\succ_j a_m$ for all $j\in N$, we have
    \begin{align*}
        \p[f(P)=a_1] & \leqslant \erm^{n\epsilon}\cdot \p[f(P)=a_m].
    \end{align*}
    Theorefore, we have $\beta^{m-1}\leqslant \erm^{n\epsilon}$, i.e., $\beta\leqslant \erm^{\frac{n\epsilon}{m-1}}$. Then Proposition~\ref{prop: tradeoff-DP-Pareto} follows.
\end{pfsketch}

Proposition \ref{prop: tradeoff-DP-Pareto} provides an upper bound on the achievable level of Pareto efficiency when subject to the constraint of $\epsilon$-DP. Next, we propose a mechanism (Mechanism~\ref{algo: BordaEXP}, Borda score exponential mechanism or BordaEXP for short) to show a lower bound of the achievable approximate Pareto efficiency under DP. Technically, Mechanism~\ref{algo: BordaEXP} is an exponential mechanism that employs the Borda score as the utility metric, where the Borda score of an alternative $a\in A$ for a given profile $P$ is defined as follows.
\begin{align*}
    \borda_P(a)=\sum\limits_{\succ_j\in P}|\{b\in A: a\succ_j b\}|.
\end{align*}

\begin{algorithm}[H]
    \caption{Borda Score Exponential Mechanism (BordaEXP)}
    \label{algo: BordaEXP}
    \KwIn{Profile $P$, Noise Level $\epsilon$}
    \KwOut{Winning Alternative}
    Get Borda score $\borda_P(a)$ of each alternative $a\in A$ for profile $P$\;
    Compute the probability distribution $p\in \Delta(A)$, such that $p(a)\propto \erm^{\borda_P(a)\epsilon/(2m-2)}$ for all $a\in A$\;
    Sample $a_{win}\sim p$\;
    \Return{$a_{win}$}
\end{algorithm}

\begin{figure}[ht]
    \centering
    \begin{minipage}{.48\linewidth}
        \label{fig: DP-Pareto}
        \includegraphics[width=\linewidth]{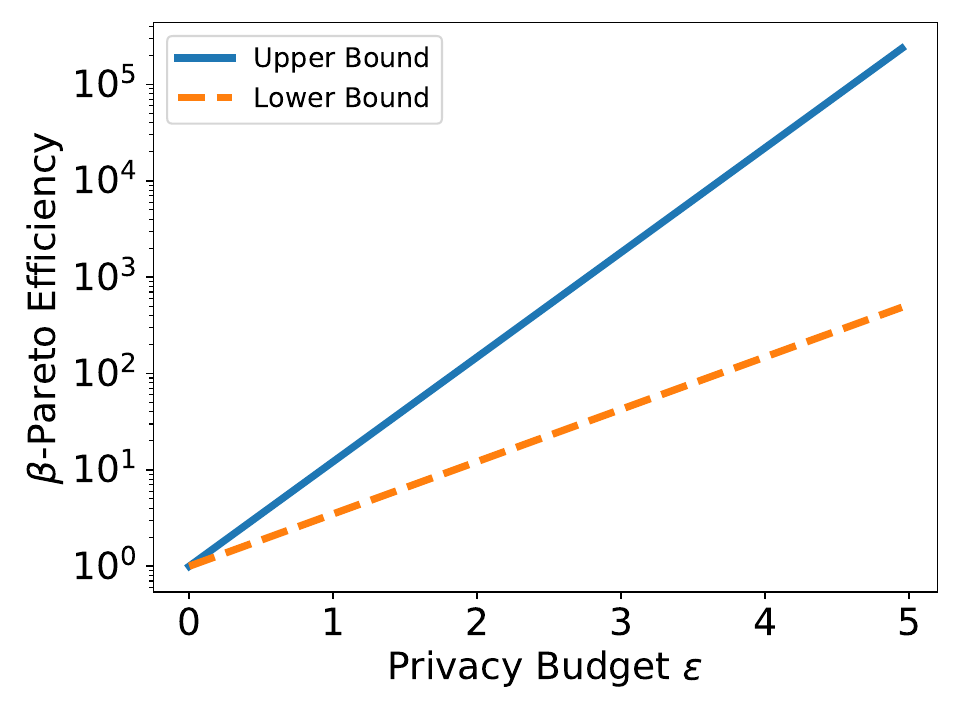}
        \caption{Upper and lower bounds of $\beta$-Pareto efficiency under $\epsilon$-DP.}
    \end{minipage}
    \ \
    \begin{minipage}{.48\linewidth}
        \label{fig: DP-SDeff}
        \includegraphics[width=\linewidth]{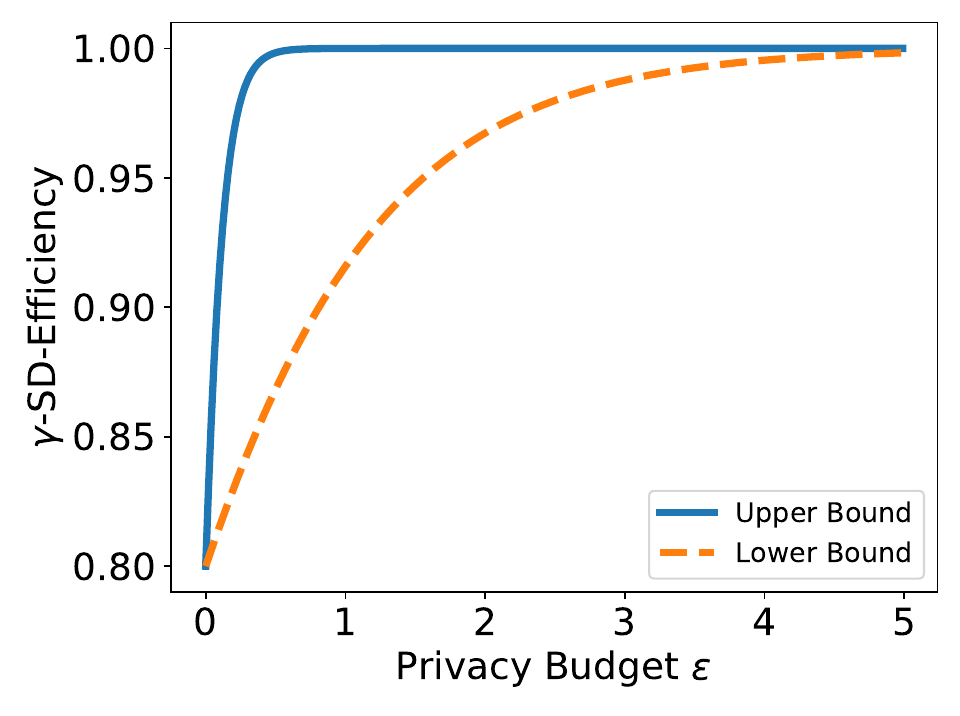}
        \caption{Upper and lower bounds of $\gamma$-SD-efficiency efficiency under $\epsilon$-DP.}
    \end{minipage}
\end{figure}

The following Proposition illustrates the lower bound of approximate Pareto efficiency achieved by Mechanism \ref{algo: BordaEXP}. 
%In fact, it is the best lower bound as far as we know. 
Both upper and lower bound for $\beta$-Pareto efficiency are $\exp(\Theta(n\epsilon/m))$. We plot both bounds in Fig. \ref{fig: DP-Pareto}, where we set $m=5, n=10$.

\begin{proposition}[\bf\boldmath $\beta$-Pareto Efficiency, Lower Bound]
    \label{prop: bordaExp-pareto}
    Given $\epsilon\in\mathbb{R}_+$, Mechanism \ref{algo: BordaEXP} satisfies $\epsilon$-DP and $\erm^{\frac{n\epsilon}{2m-2}}$-Pareto efficiency.
\end{proposition}

Secondly, we discuss the tradeoff between DP and SD-efficiency. As a notion stronger than Pareto efficiency, SD-efficiency is also incompatible with DP. In order to capture the incompatibility between DP and SD-efficiency, an approximate version of SD-efficiency is needed. Li et al.~\cite{li2022differentially} proposed approximate SD-strategyproofness, where they introduced a parameter in Inequality (\ref{equ: SD-def}) (the definition of SD relationship). Their method inspires us to consider the approximate version of SD relationship, which is defined as follows.

Let $\xi,\zeta\in \R(A)$ be two lotteries and $\succ\,\in \LA$ be a linear order on $A$. Then $\xi$ is said to $\gamma$-stochastically dominate $\zeta$ (denoted by $\xi\succeq^{\gamma-SD}\zeta$), if and only if\footnote{We use $1/\gamma$ to ensure that the property becomes stronger as $\gamma$ increases.}
\begin{align*}
    \sum\limits_{b\succ_j a} \p[\xi=b] \geqslant \frac{1}{\gamma} \cdot \sum\limits_{b\succ_j a} \p[\zeta=b].
\end{align*}
Then, we define $\gamma$-SD-efficiency as follows based on the approximate SD relationship.

\begin{definition}[\bf\boldmath $\gamma$-SD-Efficiency]
    Given $\gamma\in\mathbb{R}_+$, a voting rule $f\colon\LA^n\to\R(A)$ satisfies $\gamma$-SD-efficiency if for all profile $P$, $f(P)$ is not $\gamma$-SD-dominated.
\end{definition}

%Note that for $\gamma$-SD-efficiency, 
A larger value of $\gamma$ is %considered
more desirable  because in the definition of $\gamma$-SD, a larger $\gamma$ imposes stricter conditions for a lottery to $\gamma$-SD-dominate another one. Consequently, achieving $\gamma$-SD-efficiency becomes comparatively easier as $\gamma$ decreases. Especially, when $\gamma=1$, $\gamma$-SD-efficiency reduces to the standard SD-efficiency.

Compared to the $\beta$-Pareto efficiency (Definition \ref{def: beta-pareto}), the definition of $\gamma$-SD-efficiency is a bit more sophisticated, which brings obstacles to our exploration. Therefore, we develop a tool to help us judge whether a voting rule satisfies $\gamma$-SD-efficiency with a given $\gamma$, as shown in the following lemma.

\begin{lemma}
    \label{lem: proof-gamma-SD-eff}
    Given $\gamma>0$, a voting rule $f$ satisfies $\gamma$-SD-efficiency if and only if
    \begin{align*}
        \frac{1}{\gamma} \geqslant \sup\limits_{P\in \LA^n,\; \xi\in\R(A)}\;\inf\limits_{j\in N,\; y\in A}\frac{\sum\limits_{x:x\succ_j y} \p[\xi=x]}{\sum\limits_{x:x\succ_j y} \p[f(P)=x]}.
    \end{align*}
\end{lemma}

Using Lemma \ref{lem: proof-gamma-SD-eff}, we are ready to capture the upper bound of $\gamma$-SD-efficiency under the constraint of $\epsilon$-DP, as shown in the following proposition.

\begin{proposition}[\bf\boldmath $\gamma$-SD-Efficiency, Upper Bound]
    \label{prop: tradeoff-DP-SD-eff}
    Given $\gamma \in\mathbb{R}_+$, there is no neutral voting rule $f\colon \LA^n\to \R(A)$ satisfying $\epsilon$-DP and $\gamma$-SD-efficiency with $\gamma > \frac{(m-1)\erm^{n\epsilon}}{(m-1)\erm^{n\epsilon}+1}$.
\end{proposition}

\begin{pfsketch}
    Consider two profiles, $P_1$ and $P_2$, where all voters in $P_1$ share the same preference order $a_1\succ a_2\succ \cdots \succ a_m$. In contrast, in $P_2$, the voters' preferences are $a_m\succ' a_2\succ' a_3 \succ' \cdots\succ' a_{m-1} \succ' a_1$. Then the unique SD-efficient lottery for $P_1$ and $P_2$ should be $\mathbbm{1}_{a_1}$ and $\mathbbm{1}_{a_m}$, respectively. Here, $\mathbbm{1}_{a_1}$ and $\mathbbm{1}_{a_m}$ represent indicator functions defined as follows.
    \begin{align*}
        \p[\mathbbm{1}_{a_1}=a] = \begin{cases}
                                      1 & a = a_1          \\
                                      0 & \text{otherwise}
                                  \end{cases},\quad \p[\mathbbm{1}_{a_m}=a] = \begin{cases}
                                                                                  1 & a = a_m          \\
                                                                                  0 & \text{otherwise}
                                                                              \end{cases}.
    \end{align*}
    Let $f\colon \LA^n\to \R(A)$ be any neutral voting rule satisfying $\epsilon$-DP. Then, by Lemma \ref{lem: proof-gamma-SD-eff}, we have $\frac{1}{\gamma} \geqslant \sup\limits_{P\in \LA^n}\sup\limits_{\xi\in\R(A)}\inf\limits_{y\in A} \frac{\sum_{x\succ y} \p[\xi=x] }{\sum_{x\succ y} \p[f(P)=x] } \geqslant \frac{(m-1)\erm^{n\epsilon}+1}{(m-1)\erm^{n\epsilon}}$. Then Proposition~\ref{prop: tradeoff-DP-SD-eff} follows.
\end{pfsketch}

% Proposition 1 provides an upper bound on the achievable level of Pareto efficiency when subject to the constraint of ϵ-DP. Next, we propose a mechanism (Mechanism 1, Borda score exponential mechanism or BordaEXP for short) to show a lower bound of the achievable approximate Pareto efficiency under DP. Technically, Mechanism 1 is an exponential mechanism that employs the Borda score as the utility metric, where the Borda score of an alternative a ∈ A for a given profile P is defined as follows. 

Proposition \ref{prop: tradeoff-DP-SD-eff} illustrates an upper bound of SD-efficiency under DP. Further, based on a famous voting rule, random dictatorship\footnote{Select a voter randomly, and then announce its top-ranked alternative as the winner.} (RD), we propose a mechanism (Mechanism \ref{algo: anti-plurality-exp}, Anti-Plurality exponential RD) to explore the lower bound of the achievable approximate SD-efficiency under DP. Technically, Mechanism \ref{algo: anti-plurality-exp} replaces the dictatorial process in RD with exponential mechanism adopting anti-plurality score as its utility measure, where anti-plurality score is defined as
\begin{align*}
    \mathsf{Anti}_\succ(a)=\begin{cases}
        0, & a \text{ is the last-ranked in }\succ \\
        1, & \text{otherwise}
    \end{cases}.
\end{align*}

\begin{algorithm}[H]
    \caption{Anti-Plurality Exponential RD (RD-Anti)}
    \label{algo: anti-plurality-exp}
    \KwIn{Profile $P$, Noise Level $\epsilon$}
    \KwOut{Winning Alternative}
    \SetKwFunction{select}{Select\_Rand}
    \SetKwFunction{rand}{Rand}
    \SetKwFunction{RCM}{CM\_Rand}
    \SetKwProg{Fn}{Function}{:}{}
    Select a ballot $\succ_j\in P$ randomly\;
    Get the last-ranked alternative of $\succ_j$, denoted by $a$\;
    Compute the probability distribution $p\in\Delta(A)$, where $p(a)=\frac{1}{(m-1)\erm^\epsilon + 1}$ and $p(b)=\frac{\erm^\epsilon }{(m-1)\erm^\epsilon + 1}$, for all $b\in A\backslash\{a\}$\;
    Sample $a_{win}\sim p$\;
    \Return{$a_{win}$}
\end{algorithm}

Then the following proposition shows the lower bound of approximate SD-efficiency under $\epsilon$-DP achieved by Mechanism \ref{algo: anti-plurality-exp}. Both the upper bound and lower bound for $\gamma$-SD-efficiency are plotted in Fig. \ref{fig: DP-SDeff}, where we set $m=5, n=10$.

\begin{proposition}[\bf\boldmath $\gamma$-SD-Efficiency, Lower Bound]
    \label{prop: anti-exp-SDeff}
    Mechanism \ref{algo: anti-plurality-exp} satisfies $\epsilon$-DP and $\frac{(m-1)\erm^{\epsilon}}{(m-1)\erm^{\epsilon}+1}$-SD-efficiency.
\end{proposition}
\begin{pfsketch}
    Let $\mathfrak{E}_{\mathsf{Anti}}$ denote %the mapping introduced by 
    Mechanism \ref{algo: anti-plurality-exp}. For any neighboring profiles $P,P'\in\LA^n$ that $P_{-j}=P'_{-j}$ and $\succ_j\neq \succ_j'$, suppose that the chosen ballot in the mechanism is $\succ_i$. We define
    \begin{align*}
        C = \p[\mathfrak{E}_{\mathsf{Anti}}(P)=a~|~i\neq j] = \p[\mathfrak{E}_{\mathsf{Anti}}(P')=a~|~i\neq j] %\tag*{(use $C$ to denote them)}
    \end{align*}
    As a consequence, the DP-bound is given by the following inequality.
    \begin{align*}
        \frac{\p[\mathfrak{E}_{\mathsf{Anti}}(P)=a]}{\p[\mathfrak{E}_{\mathsf{Anti}}(P')=a]} \leqslant \frac{\erm^{\epsilon}\cdot \frac{1}{n}\p[\mathfrak{E}_{\mathsf{Anti}}(P')=a~|~i=j]+\frac{n-1}{n}\cdot C}{\frac{1}{n}\p[\mathfrak{E}_{\mathsf{Anti}}(P')=a~|~i=j]+\frac{n-1}{n}\cdot C}  \leqslant \erm^{\epsilon},
    \end{align*}
    For the bound of $\gamma$, we can prove that there is no lottery $\xi\in\R(A)$ which satisfies $\xi \succ^{\gamma-SD}_{i} \mathfrak{E}_{\mathsf{Anti}}(P)$ with $\gamma\geqslant \frac{(m-1)\erm^{\epsilon}}{(m-1)\erm^{\epsilon}+1}$ for the selected $\succ_i$, which completes the proof.
\end{pfsketch}

Finally, we investigate the tradeoff between DP and PC-efficiency. Similar to SD, the approximate version of PC relationship is also needed.

Given $\kappa>0$ and two lotteries $\xi,\zeta\in\R(A)$, $\xi$ is more desirable than $\zeta$ by means of $\kappa\text{-}PC$ (denoted by $\xi\succeq^{\kappa\text{-}PC}\zeta$) if and only if
\begin{align*}
    \sum\limits_{x\succ_i y} \p[\xi=x]\cdot \p[\zeta=y] \geqslant \frac{1}{\kappa}\cdot \sum\limits_{x\succ_i y} \p[\zeta=x]\cdot \p[\xi=y].
\end{align*}
Further, the approximate version of PC-efficiency can be defined.

\begin{definition}[\bf\boldmath $\kappa$-PC-Efficiency]
    \label{def: kappa-PC-eff}
    Given $\kappa\in\mathbb{R}_+$, a voting rule $f\colon \LA^n\to \R(A)$ satisfy $\kappa$-efficiency if $f(P)$ is never $\kappa$-PC dominated for any profile $P\in\LA^n$.
\end{definition}

Similar to the case of SD, a larger $\kappa$ in Definition \ref{def: kappa-PC-eff} is also more desirable, and $1$-PC-efficiency also reduces to standard PC-efficiency. In addition, our generalization of SD-efficiency and PC-efficiency keeps the relationship between them (PC-efficiency implies SD-efficiency), shown as follows.

\begin{lemma}
    \label{prop: PC->SD}
    $\gamma$-PC-efficiency implies $\gamma$-SD-efficiency for all $\gamma\in [0,1]$, .
\end{lemma}

Therefore, the upper bound of $\kappa$-PC-efficiency under DP must be smaller than the upper bound of $\gamma$-SD-efficiency. However, the following proposition shows that even the approximate version of PC-efficiency cannot be achieved under DP.

\begin{proposition}[\bf\boldmath $\kappa$-PC-Efficiency, Upper Bound]
    \label{prop: tradeoff-DP-PCeff}
    Given any $\kappa,\epsilon\in\mathbb{R}_+$, there is no voting rule $f\colon \LA^n\to\R(A)$ satisfying $\epsilon$-DP and $\kappa$-PC-efficiency.
\end{proposition}

\begin{pfsketch}
    Consider the profile $P$, where all voters share the same preference $a_1 \succ a_2 \succ \cdots \succ a_m$.
    Then $\mathbbm{1}_{a_1}$ can $\kappa$-PC-dominate any $f(P)$. Then, Proposition~\ref{prop: tradeoff-DP-PCeff} follows.
\end{pfsketch}

Proposition~\ref{prop: tradeoff-DP-PCeff} also implies that PC-efficiency is the limit of tradeoff between DP and efficiency. Any notion stronger than PC-efficiency cannot be even approximately achieved when DP is required.

\subsection{DP-Condorcet consistency tradeoff}

Condorcet winner and Condorcet loser are both important concepts in social choice theory. Based on them, Condorcet consistency usually refers to two axioms, i.e., Condorcet criterion and Condorcet loser criterion. However, DP requires the support set of $f(P)$ equals to $A$ for all profile $P$, i.e., $\p[f(P)=a]\neq 0$ holds for all $P\in\LA^n$ and $a\in A$, which indicates that neither of these axioms can be satisfied under the constraint of DP. Li et al.~\cite{li2022differentially} provided an approximate version of Condorcet criterion ($\alpha$-pCondorcet) to describe the level of satisfaction of Condorcet criterion. Further, they proved that if a voting rule satisfies $\epsilon$-DP and $\alpha$-pCondorcet, then $\alpha\leqslant\erm^{\epsilon}$. Such an axiom is named $\alpha$-Condorcet criterion here, of which the definition is shown as follows.

\begin{definition}[\bf\boldmath $\alpha$-Condorcet Criterion~\cite{li2022differentially}]
    Given $\alpha\in\mathbb{R}_+$, a voting rule $f\colon\LA^n\to \R(A)$ satisfies $\alpha$-Condorcet criterion if for all profiles $P$ such that $\CW(P)$ exists,
    \begin{align*}
        \p[f(P)=\CW(P)] \geqslant \alpha\cdot \p[f(P)=a],\quad \text{for all }a\in A\backslash \{\CW(P)\}.
    \end{align*}
\end{definition}

Next, we show the upper bound of approximate Condorcet criterion ($\erm^{\epsilon}$-Condorcet criterion) is achievable under DP (see Proposition~\ref{prop: CW-DP-lower}). Technically, we study Mechanism \ref{algo: CW-RR} (abbreviated as CWRR), which applies randomized response to $\mathbbm{1}_{\CW(P)}$.

\begin{algorithm}[H]
    \caption{Condorcet Winner Randomized Response (CWRR)}
    \label{algo: CW-RR}
    \KwIn{Profile $P$, Noise Level $\epsilon$}
    \KwOut{Winning Alternative}
    \If{$\CW(P)$ exists}{
        $p(\CW(P))=\frac{\erm^{\epsilon}}{\erm^\epsilon+m-1}$\;
        $p(a)=\frac{1}{\erm^{\epsilon}+m-1}$, for all $a\in A\backslash\{\CW(P)\}$\;
    }
    \Else{
        $p(a)=\frac{1}{m}$, for all $a\in A$\;
    }
    Sample $a_{win}\sim p$\;
    \Return{$a_{win}$}
\end{algorithm}

%Formally, the tightness of the upper bound $\erm^\epsilon$ is shown as follows.

\begin{proposition}[\bf\boldmath $\alpha$-Condorcet Criterion, Lower Bound]
    \label{prop: CW-DP-lower}
    Mechanism \ref{algo: CW-RR} satisfies both $\epsilon$-DP and $\erm^\epsilon$-Condorcet.
\end{proposition}

Similar to $\alpha$-Condorcet criterion, we propose the approximate Condorcet loser criterion to measure the level of satisfaction of Condorcet loser criterion.

\begin{definition}[\bf\boldmath $\eta$-Condorcet Loser Criterion]
    \label{def: eta-Condorcet-loser}
    Given $\eta\in\mathbb{R}_+$, a voting rule $f\colon\LA^n\to \R(A)$ satisfies $\eta$-Condorcet loser criterion if for all profiles $P$ such that $\CL(P)$ exists,
    \begin{align*}
        \p[f(P)=a] \geqslant \eta\cdot \p[f(P)=\CL(P)],\quad \text{for all }a\in A\backslash \{\CL(P)\}.
    \end{align*}
\end{definition}

It is notable that the larger $\eta$ is more desirable in Definition \ref{def: eta-Condorcet-loser}, and $\infty$-Condorcet loser criterion is equivalent to the standard version of Condorcet loser criterion. Further, the following proposition shows an upper bound of $\eta$ under DP.

\begin{proposition}[\bf\boldmath $\eta$-Condorcet Loser Criterion, Upper Bound]
    \label{prop: CL-DP-upper}
    There is no voting rule satisfying $\epsilon$-DP and $\eta$-Condorcet loser criterion with $\eta>\erm^{\epsilon}$.
\end{proposition}

%Intuitively, this proposition is evident, since the change on a single vote can save a Condorcet loser and shape a brand new Condorcet loser at the same time. 
Proposition~\ref{prop: CL-DP-lower} shows the upper bound in Proposition \ref{prop: CL-DP-upper} can be achieved by Mechanism \ref{algo: CL-RR} (Condorcet loser randomized response, CLRR), formally stated as follows.

\begin{algorithm}[H]
    \caption{Condorcet Loser Randomized Response (CLRR)}
    \label{algo: CL-RR}
    \KwIn{Profile $P$, Noise Level $\epsilon$}
    \KwOut{Winning Alternative}
    \If{$\CL(P)$ exists}{
        $p(\CL(P))=\frac{1}{(m-1)\erm^\epsilon+1}$\;
        $p(a)=\frac{\erm^{\epsilon}}{(m-1)\erm^\epsilon+1}$, for all $a\in A\backslash\{\CL(P)\}$\;
    }
    \Else{
        $p(a)=\frac{1}{m}$, for all $a\in A$\;
    }
    Sample $a_{win}\sim p$\;
    \Return{$a_{win}$}
\end{algorithm}

\begin{proposition}[\bf\boldmath $\eta$-Condorcet Loser Criterion, Lower Bound]
    \label{prop: CL-DP-lower}
    Mechanism \ref{algo: CL-RR} satisfies both $\erm^\epsilon$-Condorcet loser criterion and $\epsilon$-DP.
\end{proposition}

\section{Tradeoff between Axioms under DP}
\label{sec: 3-way-tradeoff}

This section investigates the 3-way tradeoffs among DP and voting axioms. Concretely, we examine the distinction between the axiom tradeoffs in classical social choice theory and the axiom tradeoffs under DP. Since even the approximation of PC-efficiency cannot be achieved under DP, we only take $\alpha$-Condorcet criterion, $\eta$-Condorcet loser criterion, $\beta$-Pareto efficiency, and $\gamma$-SD-efficiency into consideration. On the one hand, we capture the lower bounds of 3-way tradeoffs by proving the levels of satisfaction to all the approximate axioms achieved by Mechanisms \ref{algo: BordaEXP}-\ref{algo: CL-RR}, which are summarized in Table \ref{tab: 3way-tradeoff-lower}. All of the detailed results and their proofs corresponding to Table \ref{tab: 3way-tradeoff-lower} are shown in Appendix B.1.

\begin{table}
    \caption{Lower bounds of 3-way tradeoff achieved by Mechanisms \ref{algo: BordaEXP}-\ref{algo: CL-RR}.}
    \label{tab: 3way-tradeoff-lower}
    \centering
    \setlength\tabcolsep{5pt}
    \renewcommand{\arraystretch}{1.3}
    \begin{tabular}{cccccc}
        \toprule
        Voting rule & \makecell{$\beta$-Pareto                                                                                                                                                                                                                                                                                \\ efficiency} & $\gamma$-SD-efficiency & \makecell{$\alpha$-Condorcet \\ criterion} & \makecell{$\eta$-Condorcet \\ loser criterion} & Reference \\
        \midrule
        BordaEXP    & $\erm^{\frac{n\epsilon}{2m-2}}$ & $\frac{\erm^{\frac{n}{2}}+(m-2)\cdot \erm^{\frac{n(m-2)}{4m-4}}}{\erm^{\frac{n}{2}}+(m-1)\cdot \erm^{\frac{n(m-2)}{4m-4}}}$ & $\erm^{\left(\lfloor \frac{n}{2} \rfloor + 1\right)\cdot \frac{m}{2m-2}-\frac{n}{2}}$ & $\erm^{\frac{n}{2m-2}-\left( \lceil \frac{n}{2} \rceil - 1 \right)\frac{m}{2m-2}}$ & Mechanism \ref{algo: BordaEXP}           \\
        RD-Anti     & $1$                             & $\frac{(m-1)\erm^{\epsilon}}{(m-1)\erm^{\epsilon}+1}$                             & $\frac{\left(\lfloor\frac{n}{2}\rfloor-1\right)\erm^{\epsilon}+\lceil\frac{n}{2}\rceil+1}{n\erm^{\epsilon}}$                                                                     & $\frac{\left(\lfloor\frac{n}{2}\rfloor-1\right)\erm^{\epsilon}+\lceil\frac{n}{2}\rceil+1}{n\erm^{\epsilon}}$                                                          & Mechanism \ref{algo: anti-plurality-exp} \\
        CWRR        & $1$                             & $\frac{m-1}{m}$                                 & $\erm^{\epsilon}$                                                       & $1$                                                          & Mechanism \ref{algo: CW-RR}              \\
        CLRR        & $1$                             & $\frac{(m-2)\erm^{\epsilon}+1}{(m-1)\erm^{\epsilon}+1}$                             & $1$                                                                     & $\erm^{\epsilon}$                                            & Mechanism \ref{algo: CL-RR}              \\
        \bottomrule
    \end{tabular}
\end{table}

On the other hand, we investigate the upper bounds of the 3-way tradeoff between DP and each pairwise combination of the axioms. The results are listed in Table \ref{tab: 3way-tradeoff}. All of the missing proofs for the upper bounds in this section can be found in Appendix B.2.

\begin{table}
    \caption{Upper bounds of the 3-way tradeoffs among DP and axioms.}
    \label{tab: 3way-tradeoff}
    \centering
    \setlength\tabcolsep{5pt}
    \renewcommand{\arraystretch}{1.3}
    \begin{tabular}{ccc}
        \toprule
        Combination of axioms                                            & Upper bounds                                                                                                                 & Reference                                            \\
        \midrule
        $\alpha$-Condorcet criterion - $\eta$-Condorcet loser criterion~ & $\alpha\cdot\eta\leqslant \erm^{\epsilon}$                                                                                   & Theorem \ref{thm: Condorcet-winner-loser}            \\

        $\alpha$-Condorcet criterion - $\beta$-Pareto efficiency~        & $\alpha\cdot\beta^{m-2}\leqslant \erm^{n\epsilon}$                                                                           & Theorem \ref{thm: tradeoff-DP-Pareto-Condorcet}      \\

        $\eta$-Condorcet loser criterion - $\beta$-Pareto efficiency~    & $\eta\cdot\beta^{m-2}\leqslant \erm^{n\epsilon}$                                                                             & Theorem \ref{thm: tradeoff-DP-Pareto-CondorcetLoser} \\
        
        $\alpha$-Condorcet criterion - $\gamma$-SD-efficiency~           & $\gamma \leqslant \frac{\alpha+m-1-\alpha\erm^{-n\epsilon}}{\alpha+m-1}$                                                     & Theorem \ref{thm: tradeoff-DP-CW-SD}                 \\

        $\eta$-Condorcet loser criterion - $\gamma$-SD-efficiency~       & $\gamma \leqslant \frac{\erm^{n\epsilon}-\eta}{\erm^{n\epsilon}}$                                                            & Theorem \ref{thm: tradeoff-DP-CL-SD}                 \\

        $\beta$-Pareto efficiency - $\gamma$-SD-efficiency~              & $\gamma \leqslant \frac{\erm^{n\epsilon}-\erm^{n\epsilon}\beta^{2-m}}{\erm^{n\epsilon}-\erm^{n\epsilon}\beta^{2-m}+\beta-1}$ & Theorem \ref{thm: SD-PE}                             \\
        \bottomrule
    \end{tabular}
\end{table}

\parahead{Lower bounds of 3-way tradeoffs} In the previous section, we showed that BordaExp, RD-Anti, CWRR, and CLRR currently reach the best achievable levels of $\beta$-Pareto efficiency, $\gamma$-SD-efficiency, $\alpha$-Condorcet criterion, and $\eta$-Condorcet loser criterion, respectively. Therefore, these voting rules can provide the lower bounds of 3-way tradeoffs. For example, CWRR satisfies $1$-Pareto efficiency and $\erm^{\epsilon}$-Condorcet criterion, which indicates that the 3-way tradeoff among $\epsilon$-DP, $\beta$-Pareto efficiency, and $\alpha$-Condorcet criterion has a lower bound, i.e., $\alpha=\erm^{\epsilon}$ and $\beta=1$. 

Especially, the smoothed tradeoff between $\alpha$-Condorcet criterion and $\eta$-Condorcet loser criterion can be achieved by the probability mixture between CWRR (Mechanism \ref{algo: CW-RR}) and CLRR (Mechanism \ref{algo: CL-RR}), presented in Mechanism \ref{algo: CWRR-CLRR}.

\begin{algorithm}[H]
    \caption{Probability Mixture of CWRR and CLRR}
    \label{algo: CWRR-CLRR}
    \KwIn{Profile $P$, Noise Level $\epsilon$, Probability $\omega$}
    \KwOut{Winning Alternative}
    Sample $x\sim\texttt{Bernouli}(\omega)$, i.e., $\p[x=1]=\omega$ and $\p[x=0]=1-\omega$\;
    \If{$\omega=1$}{
        \Return $\text{CWRR}(P, \epsilon)$\;
    }
    \Else{
        \Return $\text{CLRR}(P,\epsilon)$\;
    }
\end{algorithm}

Let $f_\epsilon^\omega(P)$ denote $\omega\cdot \text{CWRR}(P, \epsilon)+(1-\omega)\cdot \text{CLRR}(P,\epsilon)$. Then for any profile $P$,

\begin{align*}
    \p[f_\epsilon^\omega(P)=a] = \begin{cases}
                            \frac{\omega\cdot \erm^{\epsilon}}{\erm^{\epsilon}+m-1}+\frac{(1-\omega)\cdot \erm^{\epsilon}}{(m-1)\cdot \erm^{\epsilon}+1}, & a=\CW(P)         \\
                            \frac{\omega}{\erm^{\epsilon}+m-1}+\frac{1-\omega}{(m-1)\cdot \erm^{\epsilon}+1},                                             & a=\CL(P)         \\
                            \frac{\omega}{\erm^{\epsilon}+m-1}+\frac{(1-\omega)\cdot \erm^{\epsilon}}{(m-1)\cdot \erm^{\epsilon}+1},                      & \text{otherwise}
                        \end{cases}.
\end{align*}
Therefore, $f_\epsilon^\omega$ satisfies $\epsilon$-DP, $\alpha$-Condorcet criterion, and $\eta$-Condorcet loser criterion, where
\begin{align*}
    \alpha=\frac{\frac{\omega\cdot \erm^{\epsilon}}{\erm^{\epsilon}+m-1}+\frac{(1-\omega)\cdot \erm^{\epsilon}}{(m-1)\cdot \erm^{\epsilon}+1}}{\frac{\omega}{\erm^{\epsilon}+m-1}+\frac{(1-\omega)\cdot \erm^{\epsilon}}{(m-1)\cdot \erm^{\epsilon}+1}},\quad \eta=\frac{\frac{\omega}{\erm^{\epsilon}+m-1}+\frac{(1-\omega)\cdot \erm^{\epsilon}}{(m-1)\cdot \erm^{\epsilon}+1}}{\frac{\omega}{\erm^{\epsilon}+m-1}+\frac{1-\omega}{(m-1)\cdot \erm^{\epsilon}+1}}.
\end{align*}
In other words, $\alpha\eta=\erm^{\epsilon}$. Please refer to Appendix B.1 for the full results of lower bounds. In the rest of this section, we will show the upper bounds of 3-way tradeoffs.

\parahead{Tradeoff between Condorcet consistency} First of all, we discuss the tradeoff between Condorcet criterion and Condorcet loser criterion. In fact, the standard forms of these two axioms are compatible in standard social choice theory (without DP), since for any given profile $P$, the Condorcet winner $\CW(P)$ will never coincide with the Condorcet loser $\CL(P)$. However, when DP is required, the best Condorcet criterion ($\alpha=\erm^{\epsilon}$) is not compatible with the best Condorcet loser criterion ($\eta=\erm^{\epsilon}$), since Condorcet winner can sometimes be converted to Condorcet loser by reversing only one voter's vote (e.g., when $P_{-j}$ are tied, reversing $\succ_j$ exchanges Condorcet winner and Condorcet loser). Formally, we have the following theorem.

\begin{theorem}
    \label{thm: Condorcet-winner-loser}
    There is no voting rule satisfying $\epsilon$-DP, $\alpha$-Condorcet criterion and $\eta$-Condorcet loser criterion with $\alpha\cdot \eta>\erm^{\epsilon}$.
\end{theorem}
\begin{pfsketch}
    Consider the profile $P$ ($n=2k+1$), where $k+1$ voters consider $a_1\succ a_2\succ \cdots\succ a_m$ and $k$ voters consider $a_m\succ a_{m-1}\succ\cdots\succ a_1$. Let $P'$ be another profile ($n=2k+1$), where $k$ voters consider $a_1\succ' a_2\succ' \cdots\succ' a_m$, and $k+1$ voters consider $a_m\succ a_{m-1}\succ\cdots\succ a_1$. Then $P$ and $P'$ are neighboring, and $\CW(P)=\CL(P')$. By the definition of DP, we have $\alpha\cdot\eta\leqslant \erm^{\epsilon}$.
\end{pfsketch}

\begin{figure}[ht]
    \centering
    \includegraphics[width=.5\linewidth]{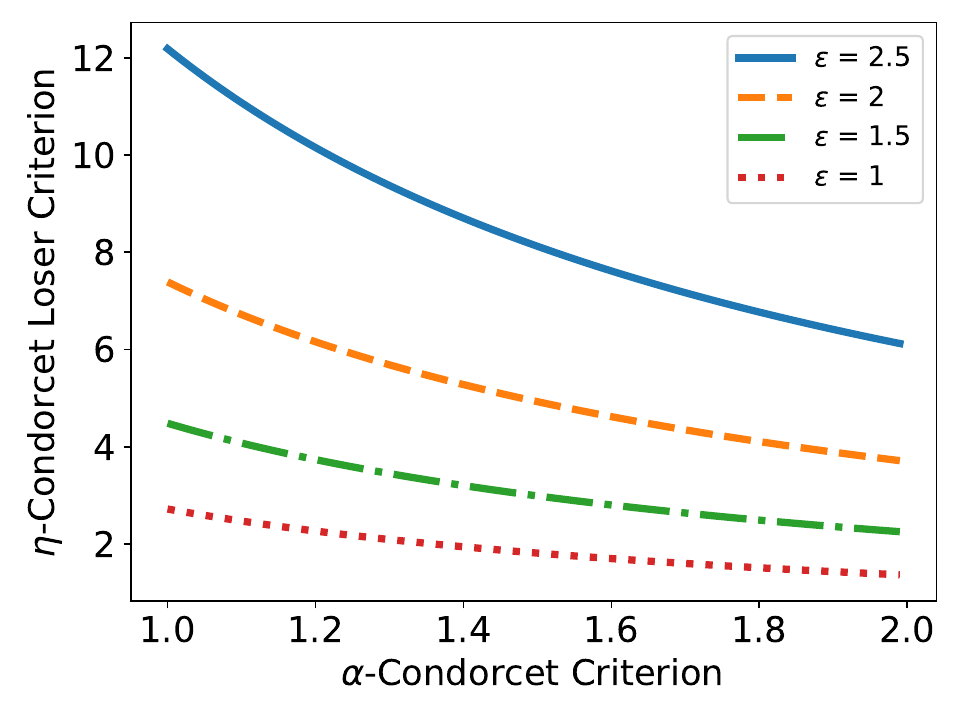}
    \caption{The Tradeoff between Condorcet criterion and Condorcet loser criterion under DP (upper bounds).}
    \label{fig: CW-CL}
\end{figure}

Surprisingly, the upper bound shown in Theorem \ref{thm: Condorcet-winner-loser} coincides with the lower bound achieved by the probability mixture of CWRR and CLRR. In other words, this bound is tight for Condorcet criterion and Condorcet loser criterion, which is visualized in Fig. \ref{fig: CW-CL}. Note that this curve is only related to $\epsilon$ and does not depend on $m$ or $n$ (number of alternatives and voters).

\parahead{Condorcet consistency against Pareto efficiency} Secondly, we discuss the 3-way tradeoff between Condorcet consistency and Pareto efficiency. In standard social choice theory, Pareto efficiency and Condorcet criterion are compatible, since selecting the original Condorcet voting method (select $\CW(P)$ as the winner with probability $1$) satisfies Pareto efficiency. However, under the constraint of $\epsilon$-DP, $\alpha$-Condorcet criterion only focuses on maximizing the winning probability of $\CW(P)$, while $\beta$-Pareto efficiency requires us to consider each pair of Pareto dominating and Pareto dominated alternatives. As a result, when DP is required, the best Pareto efficiency and the best Condorcet criterion may not be achieved simultaneously. Formally, the following proposition illustrates their tradeoff under DP.

\begin{theorem}
    \label{thm: tradeoff-DP-Pareto-Condorcet}
    There is no neutral voting rule $f\colon \LA^n\to A$ satisfying $\epsilon$-DP, $\beta$-Pareto efficiency, and $\alpha$-Condorcet criterion with $\alpha\beta^{m-2}> \erm^{n\epsilon}$.
\end{theorem}
\begin{pfsketch}
    Let $P$ be a profile ($n=2k+1$), where $k+1$ voters consider $a_1\succ a_2\succ\cdots\succ a_m$ and the rest $k$ voters consider $a_2\succ \cdots\succ a_m\succ a_1$. Then $a_1$ is the Condorcet winner, ``dominating'' any other alternatives, and each $a_i$ Pareto dominates $a_{i+1}$, which can be visualized as follows.
    \begin{align}
        \label{equ: Condorcet->Pareto}
        a_1\xrightarrow{\bf Condorcet~Winner} a_2\xrightarrow{\bf Pareto} a_3\xrightarrow{\bf Pareto} \cdots \xrightarrow{\bf Pareto} a_m.
    \end{align}
    Together with Equation (\ref{equ: n-eps-pareto-bound}), the proof can be completed.
\end{pfsketch}

According to Proposition \ref{prop: tradeoff-DP-Pareto} and~\cite{li2022differentially}, the upper bounds of Pareto efficiency and Condorcet criterion are $\erm^{\frac{n}{m-1}}$ and $\erm^\epsilon$, respectively. Therefore, there is a natural upper bound of $\alpha\beta^{m-2}$,
\begin{align}
    \label{equ: natural-bound-CW-Pareto}
    \alpha\beta^{m-2}\leqslant \sup\alpha\cdot (\sup\beta)^{m-2} = \erm^{\epsilon}\cdot \erm^{\frac{n}{m-1}\cdot(m-2)} = \erm^{1+n-\frac{n}{m-1}}.
\end{align}
Then it follows that the upper bound in Proposition \ref{thm: tradeoff-DP-Pareto-Condorcet} is better than this bound if and only if $n \leq m-1$. In fact, when $n > m-1$, Diagram (\ref{equ: Condorcet->Pareto}) does not suggest any incompatibility between Pareto efficiency and Condorcet criterion, as the upper bound illustrated in Theorem \ref{thm: tradeoff-DP-Pareto-Condorcet} is even worse than the natural bound shown in Equation (\ref{equ: natural-bound-CW-Pareto}).

Similarly, the same phenomenon also occurs to the Condorcet loser criterion. Condorcet loser criterion is also compatible with Pareto efficiency in social choice theory, since the Condorcet loser will never be a Pareto dominator. However, due to the same reason as Condorcet criterion, the best $\eta$-Condorcet loser criterion is not compatible with the best $\beta$-Pareto efficiency under the same condition. Formally, we have the following theorem, whose proof (as shown in Appendix B), is similar to Proposition \ref{thm: tradeoff-DP-Pareto-Condorcet}.

\begin{theorem}
    \label{thm: tradeoff-DP-Pareto-CondorcetLoser}
    There is no neutral voting rule $f\colon \LA^n\to \R(A)$ satisfying $\epsilon$-DP, $\beta$-Pareto efficiency, and $\alpha$-Condorcet loser criterion with $\alpha\beta^{m-2}> \erm^{n\epsilon}$.
\end{theorem}

\parahead{Condorcet consistency against SD-efficiency} The relationship between SD-efficiency and Condorcet criterion is also affected by $\epsilon$-DP. The next proposition shows SD-efficiency is compatible with Condorcet criterion in social choice theory. To simplify notations, we let Condorcet domain $\mathcal{D}_C$ denote the set of all profiles $P$ where $\CW(P)$ exists. We say a voting rule $f\colon \LA^n\to \R(A)$ is a Condorcet method if it satisfies $\p[f(P)=\CW(P)]=1$.

\begin{proposition}
    Condorcet method satisfies SD-efficiency on $\mathcal{D}_C$.
\end{proposition}
\begin{pfsketch}
Let $\mathsf{CM}$ denote Condorcet method. Given profile $P$, suppose there is a lottery $\xi\in\R(A)$ that $\xi$ can SD-dominate $\mathsf{CM}(P)$. Then, by the definition of SD, we can prove that $\xi=\mathsf{CM}(P)$, a contradiction, which completes the proof.
\end{pfsketch}

However, when DP is required, we can not achieve the best $\alpha$-Condorcet criterion and $\gamma$-SD-efficiency simultaneously. The upper bound of this tradeoff is shown in the following theorem.

\begin{theorem}
    \label{thm: tradeoff-DP-CW-SD}
    There is no neutral voting rule $f\colon \LA^n\to \R(A)$ satisfying $\epsilon$-DP, $\alpha$-Condorcet criterion, and $\gamma$-SD efficiency with $\gamma > \frac{\alpha+m-1-\alpha\erm^{-n\epsilon}}{\alpha+m-1}$.
\end{theorem}
\begin{pfsketch}
    Consider the profile $P$, where all voters' votes are exactly the same, i.e., $a_1\succ_j a_2\succ_j\cdots\succ_j a_m$, for all $j\in N$. Then $\CW(P)=a_1$, and the unique SD-efficient lottery for $P$ is $\mathbbm{1}_{a_1}$. Further, $\mathbbm{1}_{a_1}$ can $\frac{\alpha+m-1-\alpha\erm^{-n\epsilon}}{\alpha+m-1}$-dominate any $f(P)$, which completes the proof.
\end{pfsketch}

\begin{figure}[ht]
    \centering
    \begin{minipage}{.48\linewidth}
        \center
        \includegraphics[width=\linewidth]{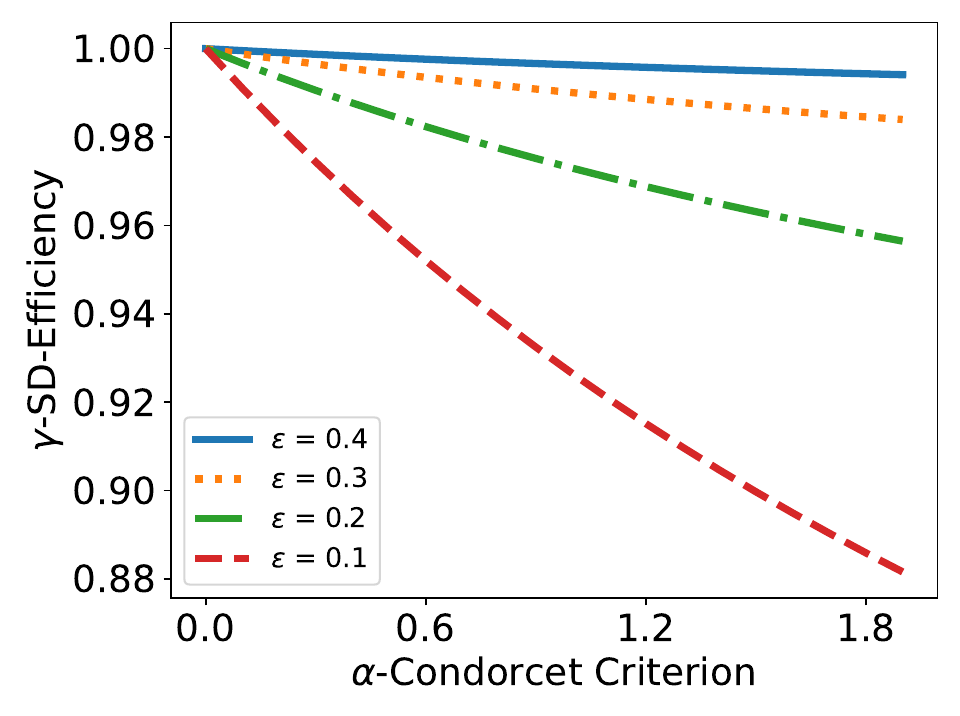}
        \caption{Tradeoff curves between $\gamma$-SD-efficiency and $\alpha$-Condorcet loser criterion (upper bounds).}
        \label{fig: CW-SDeff}
    \end{minipage}
    \ \
    \begin{minipage}{.48\linewidth}
        \center
        \includegraphics[width=\linewidth]{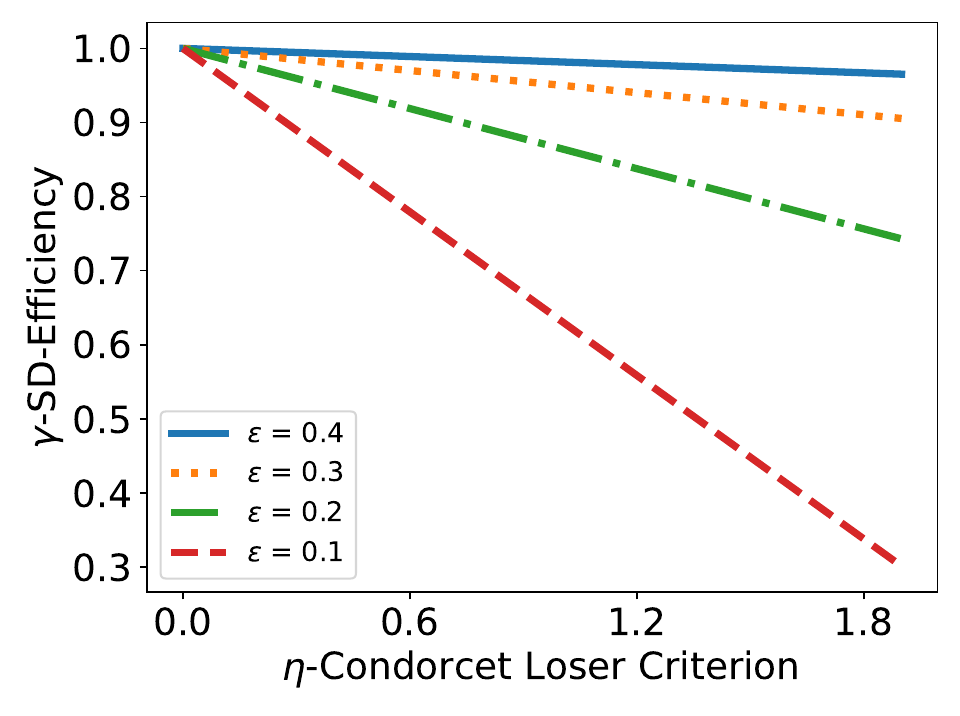}
        \caption{Tradeoff curves between
            $\gamma$-SD-efficiency and $\eta$-Condorcet loser criterion (upper bounds).}
        \label{fig: CL-SDeff}
    \end{minipage}
\end{figure}

The tradeoff curves between SD-efficiency and Condorcet criterion subject to $\epsilon$-DP are shown in Fig. \ref{fig: CW-SDeff}, where we set $m=5,n=10$.

For Condorcet loser criterion, the situation is quite similar. On the one hand, the compatibility between SD-efficiency and Condorcet loser criterion without DP is proved via the maximal-lottery mechanism~\cite{Brandt2017:Rolling}. On the other hand, there is little difference between optimizing the SD-efficiency and minimizing the winning probability of the Condorcet loser under DP, which leads to a 3-way tradeoff among them. Formally, we have the following theorem.

\begin{theorem}
    \label{thm: tradeoff-DP-CL-SD}
    There is no neutral voting rule $f\colon \LA^n\to \R(A)$ satisfying $\epsilon$-DP, $\eta$-Condorcet loser criterion, and $\gamma$-SD-efficiency with $\gamma > \frac{\erm^{n\epsilon}-\eta}{\erm^{n\epsilon}}$.
\end{theorem}
\begin{pfsketch}
    Let $m>2$. Consider the profile $P$ with $k(m-2)$ voters $(k\geqslant 1)$.
    \begin{itemize}
        \item $k$ voters: $y\succ a_1\succ a_2\succ \cdots \succ x \succ a_{m-2}$,
        \item $k$ voters: $y\succ a_2 \succ a_3\succ \cdots \succ x \succ a_1$,
        \item $k$ voters: $y\succ a_3 \succ a_4\succ \cdots \succ x \succ a_2$,
        \item $k$ voters: $\cdots$,
        \item $k$ voters: $y \succ a_{m-2} \succ a_1\succ \cdots \succ x \succ a_{m-3}$.
    \end{itemize}
    Then $\mathbbm{1}_y$ can $\frac{\erm^{n\epsilon}-\eta}{\erm^{n\epsilon}}$-SD-dominates $f(P)$, which completes the proof.
\end{pfsketch}

The tradeoff curves between SD-efficiency and Condorcet loser criterion under $\epsilon$-DP are shown in Fig. \ref{fig: CL-SDeff}, where we set $m=5$ and $n=10$.

\parahead{Pareto efficiency against SD-efficiency} Finally, we investigate the 3-way tradeoff between Pareto efficiency, SD-efficiency, and DP. Although the standard SD-efficiency implies the standard Pareto efficiency in social choice theory, their best approximate bounds are incompatible under DP. Formally, we have the following theorem.

\begin{theorem}
    \label{thm: SD-PE}
    There is no neutral voting rule $f\colon \LA^n\to \R(A)$ satisfying $\epsilon$-DP, $\gamma$-SD-efficiency, and $\beta$-Pareto efficiency with $\gamma > \frac{\erm^{n\epsilon}-\erm^{n\epsilon}\beta^{2-m}}{\erm^{n\epsilon}-\erm^{n\epsilon}\beta^{2-m}+\beta-1}$.
\end{theorem}
\begin{pfsketch}
    Consider the profile $P$, where $a_1\succ_j a_2\succ_j \cdots\succ_j a_m$, for all $j\in N$. Then we can show that any $f(P)$ will be $\frac{\erm^{n\epsilon}-\erm^{n\epsilon}\beta^{2-m}}{\erm^{n\epsilon}-\erm^{n\epsilon}\beta^{2-m}+\beta-1}$-SD-dominated by $\mathbbm{1}_{a_1}$ when $f$ satisfies $\beta$-Pareto efficiency and $\epsilon$-DP, which completes the proof.
\end{pfsketch}

\begin{figure}[H]
    \centering
    \includegraphics[width=.48\linewidth]{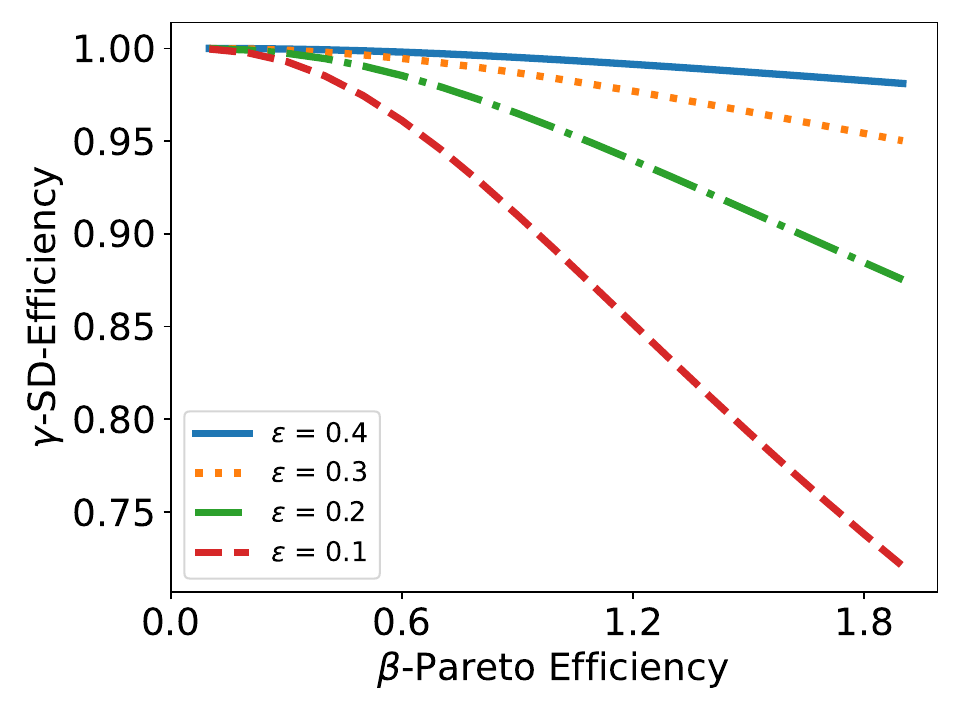}
    \includegraphics[width=.48\linewidth]{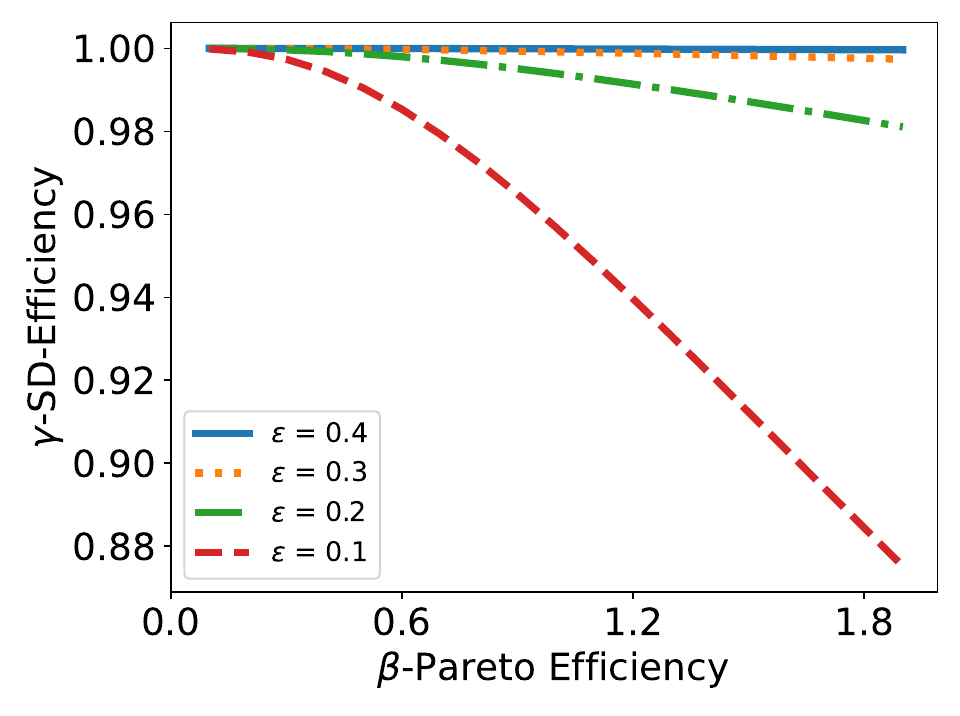}
    \caption{Tradeoff curves between $\beta$-Pareto efficiency and $\gamma$-SD-efficiency under $\epsilon$-DP (upper bounds). Left: $m=5,n=10$. Right: $m=5, n=20$.}
    \label{fig: Pareto-SDeff}
\end{figure}

The upper bounds proved in Theorem \ref{thm: SD-PE} are plotted in Fig. \ref{fig: Pareto-SDeff}. From Fig. \ref{fig: CW-CL}-\ref{fig: Pareto-SDeff}, we found the following two observations. 
\begin{enumerate}
    \item For Fig.~\ref{fig: CW-CL}-\ref{fig: Pareto-SDeff}, the curves with larger $\epsilon$ (which indicates a worse privacy guarantee) are at the top-right of the curves with smaller $\epsilon$. This indicates that the incompatibility between these axioms decreases when the privacy guarantee gets worse.
    \item For Fig.~\ref{fig: Pareto-SDeff}, the curves with larger $n$ (more voters) %(which indicates a small \textcolor{blue}{should here be larger?} number of voters) 
    are at the top-right of the curves with smaller $n$. This indicates that the incompatibility decreases as $n$ increases. The same conclusion also holds for other figures except Fig. \ref{fig: CW-CL}, where the curves are irrelevant to $n$. Figures with different $n$'s for other combinations of axioms are shown in Appendix C.
\end{enumerate}

\section{Conclusion and Future Work}
This paper investigated the tradeoff between DP and varieties of voting axioms, including Condorcet consistency and three efficiencies. We found that DP is significantly incompatible with all of these axioms and quantified their 2-way tradeoffs against DP. Further, we explored the 3-way tradeoffs among DP and these axioms. Our results show that the tradeoffs between axioms are different with or without DP. It would be an interesting future direction to study the tradeoffs between DP and other axioms. Besides, it's also interesting to develop tighter bounds for the tradeoffs.

\bibliographystyle{splncs04}
\bibliography{ref}

\begin{thebibliography}{10}
\providecommand{\url}[1]{\texttt{#1}}
\providecommand{\urlprefix}{URL }
\providecommand{\doi}[1]{https://doi.org/#1}

\bibitem{abadi2016deep}
Abadi, M., Chu, A., Goodfellow, I., McMahan, H.B., Mironov, I., Talwar, K.,
  Zhang, L.: Deep learning with differential privacy. In: Proceedings of SIGSAC
  (2016)

\bibitem{aziz2014incompatibility}
Aziz, H., Brandl, F., Brandt, F.: On the incompatibility of efficiency and
  strategyproofness in randomized social choice. In: Proceedings of AAAI (2014)

\bibitem{aziz2015universal}
Aziz, H., Brandl, F., Brandt, F.: Universal pareto dominance and welfare for
  plausible utility functions. Journal of Mathematical Economics  \textbf{60},
  123--133 (2015)

\bibitem{bassily2013coupled}
Bassily, R., Groce, A., Katz, J., Smith, A.: Coupled-worlds privacy: Exploiting
  adversarial uncertainty in statistical data privacy. In: Proceedings of FOCS
  (2013)

\bibitem{benade2019no}
Benad{\`e}, G., G{\"o}lz, P., Procaccia, A.D.: No stratification without
  representation. In: Proceedings of EC (2019)

\bibitem{berlioz2015applying}
Berlioz, A., Friedman, A., Kaafar, M.A., Boreli, R., Berkovsky, S.: Applying
  differential privacy to matrix factorization. In: Proceedings of the 9th ACM
  Conference on Recommender Systems. pp. 107--114 (2015)

\bibitem{birrell2011approximately}
Birrell, E., Pass, R.: Approximately strategy-proof voting. In: Proceedings of
  IJCAI (2011)

\bibitem{brandl2015incentives}
Brandl, F., Brandt, F., Hofbauer, J.: Incentives for participation and
  abstention in probabilistic social choice. In: Proceedings of AAMAS (2015)

\bibitem{brandl2019welfare}
Brandl, F., Brandt, F., Hofbauer, J.: Welfare maximization entices
  participation. Games and Economic Behavior  \textbf{114},  308--314 (2019)

\bibitem{DBLP:conf/ijcai/Brandl0S18}
Brandl, F., Brandt, F., Stricker, C.: An analytical and experimental comparison
  of maximal lottery schemes. In: Proceedings of IJCAI (2018)

\bibitem{Brandt2017:Rolling}
Brandt, F.: {Rolling the Dice: Recent Results in Probabilistic Social Choice}.
  In: Endriss, U. (ed.) {Trends in Computational Social Choice}, chap.~1. AI
  Access (2017)

\bibitem{Dwork06D}
Dwork, C.: Differential privacy. In: Proceedings of ICALP (2006)

\bibitem{flanigan2021fair}
Flanigan, B., G{\"o}lz, P., Gupta, A., Hennig, B., Procaccia, A.D.: Fair
  algorithms for selecting citizens’ assemblies. Nature  \textbf{596}(7873),
  548--552 (2021)

\bibitem{flanigan2020neutralizing}
Flanigan, B., G{\"o}lz, P., Gupta, A., Procaccia, A.D.: Neutralizing
  self-selection bias in sampling for sortition. In: Proceedings of NeurlPS
  (2020)

\bibitem{friedman2010data}
Friedman, A., Schuster, A.: Data mining with differential privacy. In:
  Proceedings of SIGKDD (2010)

\bibitem{gross2017vote}
Gross, S., Anshelevich, E., Xia, L.: Vote until two of you agree: Mechanisms
  with small distortion and sample complexity. In: Proceedings of AAAI (2017)

\bibitem{hay2017differentially}
Hay, M., Elagina, L., Miklau, G.: Differentially private rank aggregation. In:
  Proceedings of SDM (2017)

\bibitem{hsu2016private}
Hsu, J., Huang, Z., Roth, A., Roughgarden, T., Wu, Z.S.: Private matchings and
  allocations. SIAM Journal on Computing  \textbf{45}(6),  1953--1984 (2016)

\bibitem{kannan2018private}
Kannan, S., Morgenstern, J., Rogers, R., Roth, A.: Private pareto optimal
  exchange. ACM Transactions on Economics and Computation  \textbf{6}(3-4),
  1--25 (2018)

\bibitem{kohli2018epsilon}
Kohli, N., Laskowski, P.: Epsilon voting: Mechanism design for parameter
  selection in differential privacy. In: Proceedings of PAC (2018)

\bibitem{DBLP:conf/ijcai/Lee15}
Lee, D.T.: Efficient, private, and eps-strategyproof elicitation of tournament
  voting rules. In: Proceedings of IJCAI (2015)

\bibitem{li2020federated}
Li, T., Song, L., Fragouli, C.: Federated recommendation system via
  differential privacy. In: Proceedings of ISIT (2020)

\bibitem{li2022differentially}
Li, Z., Liu, A., Xia, L., Cao, Y., Wang, H.: Differentially private condorcet
  voting. Proceedings of AAAI  (2023)

\bibitem{ao2020private}
Liu, A., Lu, Y., Xia, L., Zikas, V.: How private are commonly-used voting
  rules? In: Proceedings of UAI (2020)

\bibitem{pai2013privacy}
Pai, M.M., Roth, A.: Privacy and mechanism design. ACM SIGecom Exchanges
  \textbf{12}(1),  8--29 (2013)

\bibitem{Plott76:Axiomatic}
Plott, C.R.: Axiomatic social choice theory: An overview and interpretation.
  American Journal of Political Science  \textbf{20}(3),  511--596 (1976)

\bibitem{procaccia2010can}
Procaccia, A.D.: Can approximation circumvent gibbard-satterthwaite? In:
  Proceedings of AAAI (2010)

\bibitem{sarwate2013signal}
Sarwate, A.D., Chaudhuri, K.: Signal processing and machine learning with
  differential privacy: Algorithms and challenges for continuous data. IEEE
  signal processing magazine  \textbf{30}(5),  86--94 (2013)

\bibitem{shang2014application}
Shang, S., Wang, T., Cuff, P., Kulkarni, S.R.: The application of differential
  privacy for rank aggregation: Privacy and accuracy. In: Proceedings of FUSION
  (2014)

\bibitem{torra2019random}
Torra, V.: Random dictatorship for privacy-preserving social choice.
  International Journal of Information Security  \textbf{19}(4), ~1--9 (2019)

\bibitem{vasa2023deep}
Vasa, J., Thakkar, A.: Deep learning: Differential privacy preservation in the
  era of big data. Journal of Computer Information Systems  \textbf{63}(3),
  608--631 (2023)

\bibitem{walsh2012lot}
Walsh, T., Xia, L.: Lot-based voting rules. In: Proceedings of AAMAS (2012)

\bibitem{wang2019aggregating}
Wang, S., Du, J., Yang, W., Diao, X., Liu, Z., Nie, Y., Huang, L., Xu, H.:
  Aggregating votes with local differential privacy: Usefulness, soundness vs.
  indistinguishability. arXiv preprint arXiv:1908.04920  (2019)

\bibitem{xiao2013privacy}
Xiao, D.: Is privacy compatible with truthfulness? In: Proceedings of ITCS
  (2013)

\bibitem{yan2020private}
Yan, Z., Li, G., Liu, J.: Private rank aggregation under local differential
  privacy. International Journal of Intelligent Systems  \textbf{35}(1),
  1492--1519 (2020)

\bibitem{zhang2011distributed}
Zhang, N., Li, M., Lou, W.: Distributed data mining with differential privacy.
  In: Proceedings of ICC (2011)

\end{thebibliography}

\newpage


\section*{Supplementary material}

# Appendix

## A  Missing Proofs in Section 3

**Proposition 1 ($\beta$-Pareto Efficiency, Upper Bound).** *There is no neutral rule $f\colon \mathcal{L}(A)^n \to \mathcal{R}(A)$ that satisfies $\epsilon$-DP and $\beta$-Pareto efficiency with $\beta > \mathrm{e}^{\frac{n\epsilon}{m-1}}$.*

*Proof.* Let $f\colon \mathcal{L}(A)^n \to A$ be a voting rule satisfying $\epsilon$-DP and $\beta$-Pareto efficiency. Let $P$ be a profile, where $a_1 \succ a_2 \succ \cdots \succ a_m$, for all $i \in N$. Since $a_1$ Pareto dominates $a_2$, $a_2$ Pareto dominates $a_3$, etc., we have

$$\mathbb{P}[f(P) = a_1] \geqslant \beta \cdot \mathbb{P}[f(P) = a_2]$$
$$\geqslant \beta^2 \cdot \mathbb{P}[f(P) = a_3]$$
$$\cdots$$
$$\geqslant \beta^{m-1} \cdot \mathbb{P}[f(P) = a_m].$$

Then we claim that for profile $P$,

$$\mathbb{P}[f(P) = a_1] \leqslant \mathrm{e}^{n\epsilon} \cdot \mathbb{P}[f(P) = a_m].$$

Thereofore, we have $\beta^{m-1} \leqslant \mathrm{e}^{n\epsilon}$, i.e., $\beta \leqslant \mathrm{e}^{\frac{n\epsilon}{m-1}}$, as desired.

Finally, we prove the claim above. In fact, for any voting rule $f\colon \mathcal{L}(A)^n \to \mathcal{R}(A)$ satisfying $\epsilon$-DP and neutrality, we have

$$\mathbb{P}[f(P) = a] \leqslant \mathrm{e}^{n\epsilon} \cdot \mathbb{P}[f(P) = b], \quad \text{for all } a, b \in A. \tag{1}$$

Now, we prove Equation (1). For any profile $P, P' \in \mathcal{L}(A)^n$, let $\ell_0(\cdot, \cdot)$ represents the $\ell_0$-distance between them, i.e., $\ell_0(P, P') = \{j \in N :\succ_j \neq \succ'_j\}$. Then, by considering the following operation **Op**, we can see that $P$ can be transferred to $P'$ through $k = \ell_0(P, P')$ times of operations.

– **Op**: Choose a voter $i \in N$ that $\succ_i \neq \succ'_i$, let $\succ_i = \succ'_i$.

Letting $P_0, P_1, \ldots, P_k$ denote all of the profiles, we have the following diagram.

$$P = P_0 \xrightarrow{\textbf{Op}} P_1 \xrightarrow{\textbf{Op}} P_2 \xrightarrow{\textbf{Op}} \cdots \xrightarrow{\textbf{Op}} P_k = P'.$$

Notice that in each step, only one voter's preference is changed. Consequently, for each $i$, $P_i$ and $P_{i+1}$ are neighboring profiles. Since $f$ satisfies $\epsilon$-DP, we have

$$\mathbb{P}[f(P) = a] \leqslant \mathrm{e}^{\epsilon} \cdot \mathbb{P}[f(P_1) = a] \leqslant \mathrm{e}^{2\epsilon} \cdot \mathbb{P}[f(P_2) = a] \leqslant \ldots \leqslant \mathrm{e}^{k \cdot \epsilon} \cdot \mathbb{P}[f(P') = a].$$

Besides, for any given $P, P' \in \mathcal{L}(A)^n$, there are at most $n$ distinct voters $j \in N$ that $\succ_j \neq \succ'_j$. Therefore, for any profile $P, P' \in \mathcal{L}(A)^n$ and any $a \in A$, we have

$$\mathbb{P}[f(P) = a] \leqslant \mathrm{e}^{n\epsilon} \cdot \mathbb{P}[f(P') = a].$$

Now, for any profile $P \in \mathcal{L}(A)^n$ and an arbitrarily chosen paif of alternatives $a, b \in A$, let $P'$ be the profile transferred from $P$ by swapping $a$ and $b$ in each voter's preference. By the neutrality of $f$, we have

$$\mathbb{P}[f(P) = a] = \mathbb{P}[f(P') = b] \text{ and } \mathbb{P}[f(P') = a] = \mathbb{P}[f(P) = b].$$

Then it follows that

$$\mathbb{P}[f(P) = a] \leqslant e^{n\epsilon} \cdot \mathbb{P}[f(P') = a]$$
$$= e^{n\epsilon} \cdot \mathbb{P}[f(P) = b],$$

which completes the proof. □

**Proposition 2 ($\beta$-Pareto Efficiency, Lower Bound).** *Given $\epsilon \in \mathbb{R}_+$, Mechanism 1 satisfies $\epsilon$-DP and $e^{\frac{n\epsilon}{2m-2}}$-Pareto efficiency.*

*Proof.* Let $\mathfrak{E}_{\mathsf{Borda}} : \mathcal{L}(A)^* \to \mathcal{R}(A)$ denote the mapping introduced by BordaEXP. Then for any profile $P \in \mathcal{L}(A)^*$ and alternative $a \in A$, we have

$$\mathbb{P}[\mathfrak{E}_{\mathsf{Borda}}(P) = a] = \frac{e^{\mathsf{Borda}_P(a)\epsilon/(2m-2)}}{\sum\limits_{c \in A} e^{\mathsf{Borda}_P(c)\epsilon/(2m-2)}}.$$

First, we establish the bound for Pareto efficiency. Given profile $P \in \mathcal{L}(A)^n$, suppose $a, b \in A$ are a pair of alternatives that $a \succ_j b$ for all $j \in N$. It follows that, for each voter $j \in N$, the number of alternatives that are considered worse than $b$ according to her preference order $\succ_j$ is strictly less than the number of alternatives considered worse than $a$. Formally, we have

$$|\{c \in A : a \succ_j c\}| - |\{c \in A : b \succ_j c\}| \geqslant 1, \quad \text{for all } j \in N.$$

By the definition of Borda score, we have

$$\mathsf{Borda}_P(a) - \mathsf{Borda}_P(b) \geqslant n.$$

Then it follows that

$$\mathbb{P}[\mathfrak{E}_{\mathsf{Borda}}(P) = a] = \frac{e^{\mathsf{Borda}_P(a) \cdot \epsilon/(2m-2)}}{\sum\limits_{c \in A} e^{\mathsf{Borda}_P(c) \cdot \epsilon/(2m-2)}}$$
$$\geqslant \frac{e^{\mathsf{Borda}_P(b)} \cdot e^{n\epsilon/(2m-2)}}{\sum\limits_{c \in A} e^{\mathsf{Borda}_P(c) \cdot \epsilon/(2m-2)}}$$
$$= e^{n\epsilon/(2m-2)} \cdot \mathbb{P}[\mathfrak{E}_{\mathsf{Borda}}(P) = b],$$

which indicates that $\mathfrak{E}_{\mathsf{Borda}}$ satisfies $e^{\frac{n\epsilon}{2m-2}}$-Pareto efficiency.

Then we prove the DP-bound. For all neighboring profiles $P, P' \in \mathcal{L}(A)^n$,

$$\frac{\mathbb{P}[\mathfrak{E}_{\mathsf{Borda}}(P) = a]}{\mathbb{P}[\mathfrak{E}_{\mathsf{Borda}}(P') = a]} = \frac{e^{\mathsf{Borda}_P(a) \cdot \frac{\epsilon}{2m-2}}}{e^{\mathsf{Borda}_{P'}(a) \cdot \frac{\epsilon}{2m-2}}} \cdot \frac{\sum\limits_{c \in A} e^{\mathsf{Borda}_{P'}(c) \cdot \frac{\epsilon}{2m-2}}}{\sum\limits_{c \in A} e^{\mathsf{Borda}_P(c) \cdot \frac{\epsilon}{2m-2}}}$$
$$\leqslant e^{\epsilon/2} \cdot \sup_{P \in \mathcal{L}(A)^n} \frac{\sum\limits_{c \in A} e^{\mathsf{Borda}_{P'}(c) \cdot \frac{\epsilon}{2m-2}}}{\sum\limits_{c \in A} e^{\mathsf{Borda}_P(c) \cdot \frac{\epsilon}{2m-2}}}$$
$$\leqslant e^{\epsilon},$$

which completes the proof. □

**Lemma 1.** *Given $\gamma > 0$, a voting rule $f$ satisfies $\gamma$-SD-efficiency if and only if*

$$\frac{1}{\gamma} \geqslant \sup_{P \in \mathcal{L}(A)^n, \xi \in \mathcal{R}(A)} \inf_{j \in N, y \in A} \frac{\sum\limits_{x:x \succ_j y} \mathbb{P}[\xi = x]}{\sum\limits_{x:x \succ_j y} \mathbb{P}[f(P) = x]}.$$

*Proof.* Suppose that $f$ is not $\gamma$-SD-efficient. Then there must be some profile $P \in \mathcal{L}(A)^n$ that $f(P)$ is $\gamma$-SD-dominated by some $\xi \in \mathcal{R}(A)$, i.e.,

$$\sum_{x:x \succ_j y} \mathbb{P}[\xi = x] \geqslant \frac{1}{\gamma} \cdot \sum_{x:x \succ_j y} \mathbb{P}[f(P) = x], \quad \text{for all } y \in A \text{ and } \succ_j \in P,$$

which is equivalent to

$$\frac{1}{\gamma} \leqslant \inf_{j \in N, y \in A} \frac{\sum\limits_{x:x \succ_j y} \mathbb{P}[\xi = x]}{\sum\limits_{x:x \succ_j y} \mathbb{P}[f(P) = x]}.$$

Therefore, $f$ is $\gamma$-SD-efficient if and only if for each $P \in \mathcal{L}(A)^n$, there does not exist such $\xi$, i.e.,

$$\frac{1}{\gamma} \geqslant \inf_{j \in N, y \in A} \frac{\sum\limits_{x:x \succ_j y} \mathbb{P}[\xi = x]}{\sum\limits_{x:x \succ_j y} \mathbb{P}[f(P) = x]}, \quad \text{for all } y \in A \text{ and } P \in \mathcal{L}(A)^n,$$

which is equivalent to

$$\frac{1}{\gamma} \geqslant \sup_{P \in \mathcal{L}(A)^n, \xi \in \mathcal{R}(A)} \inf_{j \in N, y \in A} \frac{\sum\limits_{x:x \succ_j y} \mathbb{P}[\xi = x]}{\sum\limits_{x:x \succ_j y} \mathbb{P}[f(P) = x]}.$$

That completes the proof. □

**Proposition 3 ($\gamma$-SD-Efficiency, Upper Bound).** *Given $\gamma \in \mathbb{R}_+$, there is no neutral voting rule $f \colon \mathcal{L}(A)^n \to \mathcal{R}(A)$ satisfying $\epsilon$-DP and $\gamma$-SD-efficiency with $\gamma > \frac{(m-1)e^{n\epsilon}}{(m-1)e^{n\epsilon}+1}$.*

*Proof.* Consider two profiles, $P_1$ and $P_2$, where all voters in $P_1$ share the same preference order $a_1 \succ a_2 \succ \cdots \succ a_m$. In contrast, in $P_2$, the voters' preferences are $a_m \succ' a_2 \succ' a_3 \succ' \cdots \succ' a_{m-1} \succ' a_1$. Then the unique SD-efficient lottery for $P_1$ and $P_2$ should be $\mathbb{1}_{a_1}$ and $\mathbb{1}_{a_m}$, respectively. Here, $\mathbb{1}_{a_1}$ and $\mathbb{1}_{a_m}$ represent indicator functions defined as follows.

$$\mathbb{P}[\mathbb{1}_{a_1} = a] = \begin{cases} 1 & a = a_1 \\ 0 & \text{otherwise} \end{cases}, \quad \mathbb{P}[\mathbb{1}_{a_m} = a] = \begin{cases} 1 & a = a_m \\ 0 & \text{otherwise} \end{cases}.$$

Let $f$ be any neutral voting rule satisfying $\epsilon$-DP. By Equation (1), we have

$$\mathbb{P}[f(P_1) = a_1] \leqslant e^{n\epsilon} \cdot \mathbb{P}[f(P_2) = a_1] \qquad \text{(by } \epsilon\text{-DP)}$$
$$= e^{n\epsilon} \cdot \mathbb{P}[f(P_1) = a_m]. \qquad \text{(by neutrality)}$$

By symmetry, for any $a \neq a_m$, $\mathbb{P}[f(P_1) = a] \leqslant e^{n\epsilon} \cdot \mathbb{P}[f(P_1) = a_m]$. Therefore,

$$\sum_{a \in A} \mathbb{P}[f(P_1) = a] \leqslant ((m-1)e^{n\epsilon} + 1) \cdot \mathbb{P}[f(P_1) = a_m] \leqslant 1,$$

i.e., $\mathbb{P}[f(P_1) = a_m] \leqslant \frac{1}{(m-1)\mathrm{e}^{n\epsilon}+1}$. If there exists some $\gamma$ that $f$ satisfies $\gamma$-SD-efficiency, there does not exist any $\xi \in \mathcal{R}(A)$, such that

$$\sum_{x:x \succ y} \mathbb{P}[\xi = x] \geqslant \frac{1}{\gamma} \cdot \sum_{x:x \succ y} \mathbb{P}[f(P_1) = x], \quad \text{for all } y \in A.$$

Therefore, we have

$$
\begin{aligned}
\frac{1}{\gamma} &\geqslant \sup_{P \in \mathcal{L}(A)^n} \sup_{\xi \in \mathcal{R}(A)} \inf_{y \in A} \frac{\sum\limits_{x:x \succ y} \mathbb{P}[\xi = x]}{\sum\limits_{x:x \succ y} \mathbb{P}[f(P) = x]} \\
&\geqslant \sup_{\xi \in \mathcal{R}(A)} \inf_{y \in A} \frac{\sum\limits_{x:x \succ y} \mathbb{P}[\xi = x]}{\sum\limits_{x:x \succ y} \mathbb{P}[f(P_1) = x]} \\
&\geqslant \inf_{y \in A} \frac{\sum\limits_{x:x \succ y} \mathbb{P}[\mathbb{1}_{a_1} = x]}{\sum\limits_{x:x \succ y} \mathbb{P}[f(P_1) = x]} \\
&= \frac{1}{\max\limits_{y \in A} \sum\limits_{x:x \succ y} \mathbb{P}[f(P_1) = x]} \\
&= \frac{1}{1 - \mathbb{P}[f(P_1) = a_n]} \\
&\geqslant \frac{(m-1)\mathrm{e}^{n\epsilon}+1}{(m-1)\mathrm{e}^{n\epsilon}}.
\end{aligned}
$$

In other words, we have $\gamma \leqslant \frac{(m-1)\mathrm{e}^{n\epsilon}}{(m-1)\mathrm{e}^{n\epsilon}+1}$, as desired. $\qquad\square$

**Proposition 4 ($\gamma$-SD-Efficiency, Lower Bound).** *Mechanism 2 satisfies $\epsilon$-DP and $\frac{(m-1)\mathrm{e}^{\epsilon}}{(m-1)\mathrm{e}^{\epsilon}+1}$-SD-efficiency.*

*Proof.* Letting $\mathfrak{E}_{\mathsf{Anti}} \colon \mathcal{L}(A)^n \to \mathcal{R}(A)$ denote the mapping introduced by Mechanism 2.

For any neighboring profiles $P, P' \in \mathcal{L}(A)^n$ that $P_{-j} = P'_{-j}$ and $\succ_j \neq \succ'_j$, suppose that the chosen ballot in the mechanism is $\succ_i$. Then

$$\mathbb{P}[\mathfrak{E}_{\mathsf{Anti}}(P) = a \mid i \neq j] = \mathbb{P}[\mathfrak{E}_{\mathsf{Anti}}(P') = a \mid i \neq j] \qquad \text{(use $C$ to denote them)}$$

Further, for any $a \in A$,

$$
\begin{aligned}
\frac{\mathbb{P}[\mathfrak{E}_{\mathsf{Anti}}(P) = a]}{\mathbb{P}[\mathfrak{E}_{\mathsf{Anti}}(P') = a]} &= \frac{\mathbb{P}[i = j \wedge \mathfrak{E}_{\mathsf{Anti}}(P) = a] + \mathbb{P}[i \neq j \wedge \mathfrak{E}_{\mathsf{Anti}}(P) = a]}{\mathbb{P}[i = j \wedge \mathfrak{E}_{\mathsf{Anti}}(P') = a] + \mathbb{P}[i \neq j \wedge \mathfrak{E}_{\mathsf{Anti}}(P') = a]} \\
&= \frac{\frac{1}{n}\mathbb{P}[\mathfrak{E}_{\mathsf{Anti}}(P) = a \mid i = j] + \frac{n-1}{n}\mathbb{P}[\mathfrak{E}_{\mathsf{Anti}}(P) = a \mid i \neq j]}{\frac{1}{n}\mathbb{P}[\mathfrak{E}_{\mathsf{Anti}}(P') = a \mid i = j] + \frac{n-1}{n}\mathbb{P}[\mathfrak{E}_{\mathsf{Anti}}(P') = a \mid i \neq j]} \\
&= \frac{\frac{1}{n}\mathbb{P}[\mathfrak{E}_{\mathsf{Anti}}(P) = a \mid i = j] + \frac{n-1}{n} \cdot C}{\frac{1}{n}\mathbb{P}[\mathfrak{E}_{\mathsf{Anti}}(P') = a \mid i = j] + \frac{n-1}{n} \cdot C} \qquad (P_{-j} = P'_{-j}) \\
&\leqslant \frac{\mathrm{e}^{\epsilon} \cdot \frac{1}{n}\mathbb{P}[\mathfrak{E}_{\mathsf{Anti}}(P') = a \mid i = j] + \frac{n-1}{n} \cdot C}{\frac{1}{n}\mathbb{P}[\mathfrak{E}_{\mathsf{Anti}}(P') = a \mid i = j] + \frac{n-1}{n} \cdot C} \\
&\leqslant \mathrm{e}^{\epsilon},
\end{aligned}
$$

which indicates that $\mathfrak{E}_{\mathsf{Anti}}$ satisfies $\epsilon$-DP. On the other hand, given profile $P$, suppose the top-ranked and the last-ranked alternative of $\succ_i$ are $a_{top}$ and $a_{last}$, respectively. Then, for any $\xi \in \mathcal{R}(A)$, we have

$$\frac{\sum\limits_{x:x\succ_i y}\mathbb{P}[\xi=x]}{\sum\limits_{x:x\succ_i y}\mathbb{P}[\mathfrak{E}_{\mathsf{Anti}}(P)=x]} \leqslant \frac{1}{\sum\limits_{x:x\succ_i y}\mathbb{P}[\mathfrak{E}_{\mathsf{Anti}}(P)=x]} = \frac{\sum\limits_{x:x\succ_i y}\mathbb{P}[\mathbb{1}_{a_{top}}=x]}{\sum\limits_{x:x\succ_i y}\mathbb{P}[\mathfrak{E}_{\mathsf{Anti}}(P)=x]}.$$

Thereofore,

$$\sup_{\xi\in\mathcal{R}(A)}\inf_{y\in A}\frac{\sum\limits_{x:x\succ_i y}\mathbb{P}[\xi=x]}{\sum\limits_{x:x\succ_i y}\mathbb{P}[\mathfrak{E}_{\mathsf{Anti}}(P)=x]} = \inf_{y\in A}\frac{1}{\sum\limits_{x:x\succ_i y}\mathbb{P}[\mathfrak{E}_{\mathsf{Anti}}(P)=x]}$$

$$= \frac{1}{1-\mathbb{P}[\mathfrak{E}_{\mathsf{Anti}}(P)=a_{last}]}$$

$$= \frac{(m-1)\mathrm{e}^\epsilon}{(m-1)\mathrm{e}^\epsilon+1}.$$

By Lemma 1, $\mathfrak{E}_{\mathsf{Anti}}$ satisfies $\frac{(m-1)\mathrm{e}^\epsilon}{(m-1)\mathrm{e}^\epsilon+1}$-SD-efficiency, which completes the proof. $\qquad\square$

**Lemma 2.** *Given $\gamma \leqslant 1$, $\gamma$-PC-efficiency implies $\gamma$-SD-efficiency.*

*Proof.* In fact, we only need to prove that for any $\xi, \zeta \in \mathcal{R}(A)$, $\xi \succeq^{\gamma-SD} \zeta$ implies $\xi \succeq^{\gamma-PC} \zeta$. Let $\xi$ and $\zeta$ be two lotteries satisfying $\xi \succeq^{\gamma-SD} \zeta$, i.e.,

$$\sum_{x\succ y}\mathbb{P}[\xi=x] \geqslant \frac{1}{\gamma}\cdot\sum_{x\succ y}\mathbb{P}[\zeta=y], \qquad \text{for any } y \in A.$$

Then, on the one hand, we have

$$\sum_{x,y\in A \wedge x\succ y}\mathbb{P}[\xi=x]\cdot\mathbb{P}[\zeta=y] = \sum_{y\in A}\mathbb{P}[\zeta=y]\cdot\sum_{x\succ y}\mathbb{P}[\xi=x]$$

$$\geqslant \sum_{y\in A}\mathbb{P}[\zeta=y]\cdot\frac{1}{\gamma}\sum_{x\succ y}\mathbb{P}[\zeta=y]$$

$$= \frac{1}{\gamma}\cdot\sum_{x,y\in A \wedge x\succ y}\mathbb{P}[\zeta=y]\cdot\mathbb{P}[\zeta=y].$$

On the other hand, we have

$$\frac{1}{\gamma}\cdot\sum_{x,y\in A \wedge x\succ y}\mathbb{P}[\zeta=y]\mathbb{P}[\xi=y] = \sum_{x\in A}\mathbb{P}[\zeta=y]\cdot\sum_{y\prec x}\mathbb{P}[\xi=y]$$

$$= \frac{1}{\gamma}\cdot\sum_{x\in A}\mathbb{P}[\zeta=y]\cdot\sum_{y\prec x}\left(1-\sum_{x\succeq y}\mathbb{P}[\xi=x]\right)$$

$$\leqslant \frac{1}{\gamma}\cdot\sum_{x\in A}\mathbb{P}[\zeta=y]\cdot\sum_{y\prec x}\left(1-\frac{1}{\gamma}\sum_{x\succeq y}\mathbb{P}[\zeta=y]\right)$$

$$\leqslant \frac{1}{\gamma}\cdot\sum_{x\in A}\mathbb{P}[\zeta=y]\cdot\sum_{y\prec x}\left(1-\sum_{x\succeq y}\mathbb{P}[\zeta=y]\right)$$

$$= \frac{1}{\gamma}\cdot\sum_{x,y\in A \wedge x\succ y}\mathbb{P}[\zeta=y]\cdot\mathbb{P}[\zeta=y].$$

Then it follows that

$$\sum_{x,y \in A \land x \succ y} \mathbb{P}[\xi = x] \cdot \mathbb{P}[\zeta = y] \geqslant \frac{1}{\gamma} \cdot \sum_{x,y \in A \land x \succ y} \mathbb{P}[\zeta = x] \cdot \mathbb{P}[\xi = y],$$

which completes the proof. $\qquad\square$

**Proposition 5 ($\kappa$-PC-Efficiency, Upper Bound).** *Given any $\kappa, \epsilon \in \mathbb{R}_+$, there is no voting rule $f \colon \mathcal{L}(A)^n \to \mathcal{R}(A)$ satisfying $\epsilon$-DP and $\kappa$-PC-efficiency.*

*Proof.* Consider the profile $P$, where all voters share the same prefereoce

$$a_1 \succ a_2 \succ \cdots \succ a_m.$$

Then the unique PC-efficient distribution on $A$ is $\mathbb{1}_{a_1}$. Further, we have

$$\sum_{x,y \colon x \succ_i y} \mathbb{P}[\mathbb{1}_{a_1} = x] \cdot \mathbb{P}[f(P) = y] = \sum_{y \colon a_1 \succ_i y} \mathbb{P}[f(P) = y] = 1 - \mathbb{P}[f(P) = a_1].$$

However,

$$\sum_{x,y \colon x \succ_i y} \mathbb{P}[f(P) = x] \cdot \mathbb{P}[\mathbb{1}_{a_1} = y] = \sum_{x \in A} \mathbb{P}[f(P) = x] \cdot \sum_{y \colon x \succ_i y} \mathbb{P}[\mathbb{1}_{a_1} = y] = 0.$$

In other words, for all $\kappa \in \mathbb{R}_+$, the lottery $\mathbb{1}_{a_1}$ can $\kappa$-PC-dominate any $f(P)$, which completes the proof. $\qquad\square$

**Proposition 6 ($\alpha$-Condorcet Criterion, Lower Bound).** *Mechanism 3 satisfies $e^\epsilon$-Condorcet criterion and $\epsilon$-DP.*

*Proof.* Let $\mathfrak{R}_{\mathrm{CW}} \colon \mathcal{L}(A)^* \to \mathcal{R}(A)$ denote the mapping introduced by CWRR. Then for any profile $P \in \mathcal{L}(A)^*$ and alternative $a \in A$, we have

$$\mathbb{P}[\mathfrak{R}_{\mathrm{CW}}(P) = a] = \begin{cases} \frac{e^\epsilon}{e^\epsilon + m - 1}, & a = \mathrm{CW}(P) \\ \frac{1}{e^\epsilon + m - 1}, & \text{otherwise} \end{cases}, \quad \text{for all } P \text{ that } \mathrm{CW}(P) \text{ exists.}$$

By definition, it is not hard to see that CWRR satisfies $e^\epsilon$-Condorcet criterion. Thus, we only need to prove that CWRR satisfies $\epsilon$-DP. In fact, for any neighboring profiles $P, P' \in \mathcal{L}(A)^n$ and $a \in A$,

$$\begin{aligned} \frac{\mathbb{P}[\mathfrak{R}_{\mathrm{CW}}(P) = a]}{\mathbb{P}[\mathfrak{R}_{\mathrm{CW}}(P') = a]} &\leqslant \frac{\max_{a \in A} \mathbb{P}[\mathfrak{R}_{\mathrm{CW}}(P) = a]}{\max_{a \in A} \mathbb{P}[\mathfrak{R}_{\mathrm{CW}}(P') = a]} \\ &\leqslant \frac{e^\epsilon}{e^\epsilon + m - 1} \Big/ \frac{1}{e^\epsilon + m - 1} \\ &= e^\epsilon, \end{aligned}$$

which completes the proof. $\qquad\square$

**Proposition 7 ($\eta$-Condorcet Loser Criterion, Upper Bound).** *There is no voting rule satisfying $\epsilon$-DP and $\eta$-Condorcet loser criterion with $\eta > e^\epsilon$.*

*Proof.* Suppose $f \colon \mathcal{L}(A)^n \to A$ be a voting rule satisfying $\epsilon$-DP and $\eta$-Condorcet loser criterion. Consider the profile $P$ ($n = 2k + 1$):

- $k + 1$ voters: $a_1 \succ a_2 \succ \cdots \succ a_m$,

&ndash; $k$ voters: $a_m \succ a_{m-1} \succ \cdots \succ a_1$.

By definition, we have $w_P[a_m, a_i] = -1$, for all $a_i \in A \backslash \{a_m\}$, i.e., $a_m$ is the Condorcet loser. Now, letting one voter change her preferece from $a_1 \succ a_2 \succ \cdots \succ a_m$ to $a_m \succ a_{m-1} \succ \cdots \succ a_1$, we can obtain another profile $P'$:

&ndash; $k$ voters: $a_1 \succ' a_2 \succ' \cdots \succ' a_m$,
&ndash; $k+1$ voters: $a_m \succ' a_{m-1} \succ \cdots \succ' a_1$.

Now we have $w_{P'}[a_1, a_i] = -1$, for all $a_i \in A \backslash \{a\}$, i.e., $a_1$ is the Condorcet loser for $P'$. Then

$$\mathbb{P}[f(P) = a_1] \geqslant \eta \cdot \mathbb{P}[f(P) = a_m] \qquad \text{(By } \eta\text{-Condorcet loser)}$$
$$\geqslant \mathrm{e}^{-\epsilon} \cdot \eta \cdot \mathbb{P}[f(P') = a_m] \qquad \text{(By } \epsilon\text{-DP)}$$
$$\geqslant \mathrm{e}^{-\epsilon} \cdot \eta^2 \cdot \mathbb{P}[f(P') = a_1] \qquad \text{(By } \eta\text{-Condorcet loser)}$$
$$\geqslant \mathrm{e}^{-2\epsilon} \cdot \eta^2 \cdot \mathbb{P}[f(P) = a_1], \qquad (\epsilon\text{-DP})$$

which indicates that $\mathrm{e}^{-2\epsilon} \cdot \eta^2 \leqslant 1$, i.e., $\eta \leqslant \mathrm{e}^{\epsilon}$. That completes the proof. $\qquad \square$

**Proposition 8 ($\eta$-Condorcet Loser Criterion, Lower Bound).** *Mechanism 4 satisfies* $\mathrm{e}^{\epsilon}$*-Condorcet loser criterion and $\epsilon$-DP.*

*Proof.* Let $\mathfrak{R}_{\mathrm{CL}} : \mathcal{L}(A)^* \to \mathcal{R}(A)$ denote the mapping introduced by CLRR, we have

$$\mathbb{P}[\mathfrak{R}_{\mathrm{CL}}(P) = a] = \begin{cases} \frac{1}{(m-1)\mathrm{e}^{\epsilon}+1}, & a = \mathrm{CL}(P) \\ \frac{\mathrm{e}^{\epsilon}}{(m-1)\mathrm{e}^{\epsilon}+1}, & \text{otherwise} \end{cases}, \quad \text{for all } P \text{ that } \mathrm{CL}(P) \text{ exists.}$$

By definition, it is not hard to see that CLRR satisfies $\mathrm{e}^{\epsilon}$-Condorcet criterion. Thus, we only need to prove that CLRR satisfies $\epsilon$-DP. In fact, for any neighboring profiles $P, P' \in \mathcal{L}(A)^n$ and $a \in A$,

$$\frac{\mathbb{P}[\mathfrak{R}_{\mathrm{CL}}(P) = a]}{\mathbb{P}[\mathfrak{R}_{\mathrm{CL}}(P') = a]} \leqslant \frac{\max_{a \in A} \mathbb{P}[\mathfrak{R}_{\mathrm{CL}}(P) = a]}{\max_{a \in A} \mathbb{P}[\mathfrak{R}_{\mathrm{CL}}(P') = a]}$$
$$\leqslant \frac{\mathrm{e}^{\epsilon}}{(m-1)\mathrm{e}^{\epsilon}+1} \Big/ \frac{1}{(m-1)\mathrm{e}^{\epsilon}+1}$$
$$= \mathrm{e}^{\epsilon},$$

which completes the proof. $\qquad \square$

# B Missing Proofs in Section 4

## B.1 Results in Table 3 and their proofs

**Proposition 9.** *Given $\epsilon \in \mathbb{R}_+$, BordaEXP satisfies*

*(1)* $\frac{e^{\frac{n}{2}} + (m-2) \cdot e^{\frac{n(m-2)}{4m-4}}}{e^{\frac{n}{2}} + (m-1) \cdot e^{\frac{n(m-2)}{4m-4}}}$*-SD-efficiency,*

*(2)* $e^{\left(\lfloor \frac{n}{2} \rfloor + 1\right) \cdot \frac{m}{2m-2} - \frac{n}{2}}$*-Condorcet criterion,*

*(3)* $e^{\frac{n}{2m-2} - \left(\lceil \frac{n}{2} \rceil - 1\right)\frac{m}{2m-2}}$*-Condorcet loser criterion.*

*Proof.* Let $\mathfrak{E}_{\mathsf{Borda}}$ denote the voting rule introduced by BordaEXP. First, we prove (1). In fact,

$$\sup_{P,\xi} \inf_{j,y} \frac{\sum\limits_{x \succ_j y} \mathbb{P}[\xi = x]}{\sum\limits_{x \succ_j y} \mathbb{P}[\mathfrak{E}_{\mathsf{Borda}}(P) = x]} \leqslant \sup_{P} \inf_{j,y} \frac{1}{\sum\limits_{x \succ_j y} \mathbb{P}[\mathfrak{E}_{\mathsf{Borda}}(P) = x]}$$

$$= \sup_{P} \inf_{j} \frac{1}{1 - \mathbb{P}[\mathfrak{E}_{\mathsf{Borda}}(P) = a_{last}^j]}$$

$$\leqslant \frac{1}{1 - \sup_{P} \inf_{j} \mathbb{P}[\mathfrak{E}_{\mathsf{Borda}}(P) = a_{last}^j]}.$$

where $a_{last}^j$ denote the last-ranked alternative in $\succ_j$. By syemmetry, we have

$$\sup_{P} \inf_{j} \mathbb{P}[\mathfrak{E}_{\mathsf{Borda}}(P) = a_{last}^j] = \frac{e^{\frac{n(m-2)}{4m-4}}}{e^{\frac{n}{2}} + (m-1) \cdot e^{\frac{n(m-2)}{4m-4}}}.$$

Then BordaEXP satisfies $\frac{e^{\frac{n}{2}} + (m-2) \cdot e^{\frac{n(m-2)}{4m-4}}}{e^{\frac{n}{2}} + (m-1) \cdot e^{\frac{n(m-2)}{4m-4}}}$. Second, we prove (2). By definition, for any profile $P$ that $\mathrm{CW}(P)$ exists, $\mathrm{CW}(P)$ must defeat each alternative $a \neq \{\mathrm{CW}(P)\}$ in at least half of the votes, i.e., $\mathsf{Borda}_P(\mathrm{CW}(P)) \geqslant (m-1)\left(\lfloor \frac{n}{2} \rfloor + 1\right)$. And for each $a \neq \mathrm{CW}(P)$, $\mathsf{Borda}_P(a) \leqslant (m-1)n - (\lfloor \frac{n}{2} \rfloor + 1)$. Therefore,

$$\frac{\mathbb{P}[\mathfrak{E}_{\mathsf{Borda}}(P) = \mathrm{CW}(P)]}{\mathbb{P}[\mathfrak{E}_{\mathsf{Borda}}(P) = a]} \geqslant e^{\frac{(m-1)\left(\lfloor \frac{n}{2} \rfloor + 1\right)}{2m-2} - \frac{(m-1)n - (\lfloor \frac{n}{2} \rfloor + 1)}{2m-2}}$$

$$= e^{\left(\lfloor \frac{n}{2} \rfloor + 1\right) \cdot \frac{m}{2m-2} - \frac{n}{2}},$$

which indicates that BordaEXP satisfies $e^{\left(\lfloor \frac{n}{2} \rfloor + 1\right) \cdot \frac{m}{2m-2} - \frac{n}{2}}$-Condorcet criterion. Finally, we prove (3). By definition, for any profile $P$ that $\mathrm{CL}(P)$ exists, $a \neq \mathrm{CL}(P)$ must be ranked than $\mathrm{CL}(P)$ in at least a half of votes, i.e., $\mathsf{Borda}_P(\mathrm{CL}(P)) \leqslant (m-1)\left(\lfloor \frac{n}{2} \rfloor - 1\right)$. And for each $a \neq \mathrm{CL}(P)$, $\mathsf{Borda}_P(a) \geqslant n - \lceil \frac{n}{2} \rceil + 1$. Therefore,

$$\frac{\mathbb{P}[\mathfrak{E}_{\mathsf{Borda}}(P) = a]}{\mathbb{P}[\mathfrak{E}_{\mathsf{Borda}}(P) = \mathrm{CL}(P)]} \geqslant e^{\frac{n - \lceil \frac{n}{2} \rceil + 1}{2m-2} - \frac{(m-1)\left(\lfloor \frac{n}{2} \rfloor - 1\right)}{2m-2}}$$

$$= e^{\frac{n}{2m-2} - \left(\lceil \frac{n}{2} \rceil - 1\right)\frac{m}{2m-2}},$$

which indicates that BordaEXP satisfies $e^{\frac{n}{2m-2} - \left(\lceil \frac{n}{2} \rceil - 1\right)\frac{m}{2m-2}}$-Condorcet loser criterion. □

**Proposition 10.** *Given $\epsilon \in \mathbb{R}_+$, RD-Anti satisfies*

(1) $1$-*Pareto efficiency*,

(2) $\frac{\left(\lfloor \frac{n}{2} \rfloor - 1\right) e^{\epsilon} + \lceil \frac{n}{2} \rceil + 1}{n e^{\epsilon}}$-*Condorcet criterion*,

(3) $\frac{\left(\lfloor \frac{n}{2} \rfloor - 1\right) e^{\epsilon} + \lceil \frac{n}{2} \rceil + 1}{n e^{\epsilon}}$-*Condorcet loser criterion*.

*Proof.* First, given profile $P$, for any $a, b \in A$, a Pareto dominates $b$ means that $a \succ_j b$ for all $j \in N$. Then $a$, the Pareto dominator, is never ranked last in any $\succ_j$. Therefore, $\mathbb{P}[\mathfrak{E}_{\mathsf{Anti}}(P) = a] \geqslant \mathbb{P}[\mathfrak{E}_{\mathsf{Anti}}(P) = b]$, which completes the proof of (1). Second, we prove (2). For any profile $P \in \mathcal{L}(A)^n$,

$$|\{j \in N : a^j_{last} = \mathrm{CW}(P)\}| \leqslant \lceil \frac{n}{2} \rceil - 1,$$

otherwise $\mathrm{CW}(P)$ will be the Condorcet loser. Therefore,

$$\mathbb{P}[\mathfrak{E}_{\mathsf{Anti}}(P) = \mathrm{CW}(P)] \geqslant \frac{\lceil \frac{n}{2} \rceil - 1}{n} \cdot \frac{1}{(m-1)e^{\epsilon} + 1} + \frac{\lfloor \frac{n}{2} \rfloor + 1}{n} \cdot \frac{e^{\epsilon}}{(m-1)e^{\epsilon} + 1}.$$

For any $a \neq \mathrm{CW}(P)$,

$$\mathbb{P}[\mathfrak{E}_{\mathsf{Anti}}(P) = a] \leqslant \frac{e^{\epsilon}}{(m-1)e^{\epsilon} + 1}.$$

Hence, we have

$$\frac{\mathbb{P}[\mathfrak{E}_{\mathsf{Anti}}(P) = \mathrm{CW}(P)]}{\mathbb{P}[\mathfrak{E}_{\mathsf{Anti}}(P) = a]} \geqslant \frac{\left(\lfloor \frac{n}{2} \rfloor - 1\right) e^{\epsilon} + \lceil \frac{n}{2} \rceil + 1}{n e^{\epsilon}},$$

which completes the proof. Finally, we prove (3). Given a profile $P$ that $\mathrm{CL}(P)$ exists,

$$|\{j \in N : a^j_{last} = a\}| \leqslant \lceil \frac{n}{2} \rceil - 1,$$

otherwise $a$ will be the Condorcet loser. Therefore,

$$\mathbb{P}[\mathfrak{E}_{\mathsf{Anti}}(P) = a] \geqslant \frac{\lceil \frac{n}{2} \rceil - 1}{n} \cdot \frac{1}{(m-1)e^{\epsilon} + 1} + \frac{\lfloor \frac{n}{2} \rfloor + 1}{n} \cdot \frac{e^{\epsilon}}{(m-1)e^{\epsilon} + 1}.$$

Hoewver,

$$\mathbb{P}[\mathfrak{E}_{\mathsf{Anti}}(P) = \mathrm{CL}(P)] \leqslant \frac{e^{\epsilon}}{(m-1)e^{\epsilon} + 1}.$$

Hence we have

$$\frac{\mathbb{P}[\mathfrak{E}_{\mathsf{Anti}}(P) = a]}{\mathbb{P}[\mathfrak{E}_{\mathsf{Anti}}(P) = \mathrm{CL}(P)]} \geqslant \frac{\left(\lfloor \frac{n}{2} \rfloor - 1\right) e^{\epsilon} + \lceil \frac{n}{2} \rceil + 1}{n e^{\epsilon}},$$

which completes the proof of (3). □

**Proposition 11.** *Given $\epsilon \in \mathbb{R}_+$, CWRR satisfies $1$-Pareto efficiency, $\frac{m-1}{m}$-SD-efficiency, and $1$-Condorcet loser criterion.*

*Proof.* The bounds of Pareto efficiency and Condorcet loser criterion are evident, since for any profile $P$, neither a Pareto dominated alternative nor the Condorcet loser can be the Condorcet winner. Then we only need to prove the bound of SD-efficiency. Given profile $P$, we have

$$\sup_{P, \xi} \inf_{j, y} \frac{\sum_{x \succ_j y} \mathbb{P}[\xi = x]}{\sum_{x \succ_j y} \mathbb{P}[\mathfrak{R}_{\mathrm{CW}}(P) = x]} \leqslant \frac{1}{1 - \sup_P \inf_j \mathbb{P}[\mathfrak{R}_{\mathrm{CW}}(P) = a^j_{last}]}.$$

Then there are two possible cases for the profile, discussed as follows

1. If $\mathrm{CW}(P)$ exists, then there must exist some $j$ that $a_{last}^j \neq \mathrm{CW}(P)$. Therefore

$$\inf_j \mathbb{P}[\mathfrak{R}_{\mathrm{CW}}(P) = a_{last}^j] = \frac{1}{\mathrm{e}^\epsilon + m - 1}.$$

2. If $\mathrm{CW}(P)$ does not exist, then

$$\inf_j \mathbb{P}[\mathfrak{R}_{\mathrm{CW}}(P) = a_{last}^j] = \frac{1}{m} \geqslant \frac{1}{\mathrm{e}^{\epsilon+m-1}}.$$

In other words, we have

$$\sup_P \inf_j \mathbb{P}[\mathfrak{R}_{\mathrm{CW}}(P) = a_{last}^j] = \frac{1}{m},$$

which indicates that CWRR satisfies $\frac{m-1}{m}$-SD-efficiency. That completes the proof. □

**Proposition 12.** *Given $\epsilon \in \mathbb{R}_+$, CLRR satisfies 1-Pareto efficiency, $\frac{(m-2)\mathrm{e}^\epsilon+1}{(m-1)\mathrm{e}^\epsilon+1}$-SD-efficiency, and 1-Condorcet criterion.*

*Proof.* The bounds of Pareto efficiency and Condorcet criterion are evident, since for any profile $P$, neither a Pareto dominator nor the Condorcet winner can be the Condorcet loser. Then we only need to prove the bound of SD-efficiency. Given profile $P$, we have

$$\sup_{P,\xi} \inf_{j,y} \frac{\sum\limits_{x \succ_j y} \mathbb{P}[\xi = x]}{\sum\limits_{x \succ_j y} \mathbb{P}[\mathfrak{R}_{\mathrm{CL}}(P) = x]} \leqslant \frac{1}{1 - \sup\limits_P \inf\limits_j \mathbb{P}[\mathfrak{R}_{\mathrm{CL}}(P) = a_{last}^j]}.$$

Then there are two possible cases for the profile, discussed as follows

1. If $\mathrm{CL}(P)$ exists, considering the profile $P$, where each $a_{last}^j \neq \mathrm{CL}(P)$ for each $j \in N$, we have

$$\inf_j \mathbb{P}[\mathfrak{R}_{\mathrm{CL}}(P) = a_{last}^j] = \frac{\mathrm{e}^\epsilon}{(m-1)\mathrm{e}^\epsilon + 1}.$$

2. If $\mathrm{CL}(P)$ does not exist, then

$$\inf_j \mathbb{P}[\mathfrak{R}_{\mathrm{CL}}(P) = a_{last}^j] = \frac{1}{m} \leqslant \frac{\mathrm{e}^\epsilon}{(m-1)\mathrm{e}^\epsilon + 1}.$$

In other words, we have

$$\sup_P \inf_j \mathbb{P}[\mathfrak{R}_{\mathrm{CL}}(P) = a_{last}^j] = \frac{1}{m},$$

which indicates that CWRR satisfies $\frac{(m-2)\mathrm{e}^\epsilon+1}{(m-1)\mathrm{e}^\epsilon+1}$-SD-efficiency. That completes the proof. □

### B.2 Proofs of Theorems 1-6

**Theorem 1.** *There is no voting rule satisfying $\epsilon$-DP, $\alpha$-Condorcet criterion and $\eta$-Condorcet loser criterion with $\alpha \cdot \eta > \mathrm{e}^\epsilon$.*

*Proof.* Consider the profile $P$ ($n = 2k+1$):

– $k + 1$ voters: $a_1 \succ a_2 \succ \cdots \succ a_m$,

– $k$ voters: $a_m \succ a_{m-1} \succ \cdots \succ a_1$.

By definition, we have $CW(P) = a_1$, since $w_P[a_1, a_i] = 1$, for all $a_i \neq a_1$. Now consider another profile $P'$ with the same number of voters:

– $k$ voters: $a_1 \succ' a_2 \succ' \cdots \succ' a_m$,
– $k+1$ voters: $a_m \succ a_{m-1} \succ \cdots \succ a_1$.

Then $w_P[a_1, a_i] = -1$, for all $a_i \in A \backslash \{a\}$, i.e., $a_1$ is a Condorcet loser. Since there is only one voter changes her preference from $P$ to $P'$, we have

$$
\begin{aligned}
\mathbb{P}[f(P) = a_1] &\geqslant \alpha \cdot \mathbb{P}[f(P) = a_2] && (\alpha\text{-Condorcet criterion}) \\
&\geqslant \alpha\eta \cdot \mathbb{P}[f(P) = a_m] && (\eta\text{-Condorcet loser criterion}) \\
&\geqslant e^{-\epsilon} \cdot \alpha\eta \cdot \mathbb{P}[f(P') = a_m] && (\epsilon\text{-DP}) \\
&\geqslant e^{-\epsilon} \cdot \alpha^2 \cdot \eta \cdot \mathbb{P}[f(P') = a_2] && (\alpha\text{-Condorcet criterion}) \\
&\geqslant e^{-\epsilon} \cdot \alpha^2 \cdot \eta^2 \cdot \mathbb{P}[f(P') = a_1] && (\eta\text{-Condorcet loser criterion}) \\
&\geqslant e^{-2\epsilon} \cdot \alpha^2 \cdot \eta^2 \cdot \mathbb{P}[f(P') = a_1], && (\epsilon\text{-DP})
\end{aligned}
$$

which indicates that $e^{-2\epsilon}\alpha^2\eta^2 \leqslant 1$, i.e., $\alpha\eta \leqslant e^{\epsilon}$. That completes the proof. □

**Theorem 2.** *If a neutral voting rule $f \colon \mathcal{L}(A)^n \to A$ satisfies $\epsilon$-DP, $\beta$-Pareto efficiency, and $\alpha$-Condorcet criterion, then $\alpha\beta^{m-2} \leqslant e^{n\epsilon}$.*

*Proof.* Consider the following profile $P$ ($n = 2k+1$):

– $k+1$ voters: $a_1 \succ a_2 \succ \cdots \succ a_m$;
– $k$ voters: $a_2 \succ \cdots \succ a_m \succ a_1$.

By definition, we have $w_P[a_1, a_i] = 1$, for all $a_i \in A \backslash \{a_1\}$, which indicates that $CW(P) = a_1$. Also notice that $w_P[a_i, a_j] = n$ for all $i < j$. Thus, $a_i$ Pareto dominates $a_j$ for all $i < j$. The relations among all alternatives are shown in the following graph.

$$ a_1 \xrightarrow{\textbf{Condorcet Winner}} a_2 \xrightarrow{\textbf{Pareto}} a_3 \xrightarrow{\textbf{Pareto}} \cdots \xrightarrow{\textbf{Pareto}} a_m. \tag{2} $$

Since $f$ satisfies $\alpha$-Condorcet criterion and $\beta$-Pareto efficiency, we have

$$
\begin{aligned}
\mathbb{P}[f(P) = a_1] &\geqslant \alpha \cdot \mathbb{P}[f(P) = a_2] \\
&\geqslant \alpha\beta \cdot \mathbb{P}[f(P) = a_3] \\
&\geqslant \cdots \\
&\geqslant \alpha\beta^{m-2} \cdot \mathbb{P}[f(P) = a_m].
\end{aligned}
$$

Now, consider another profile $P'$, where all voters' preferences are exactly the same:

$$ a_m \succ a_{m-1} \succ \cdots \succ a_1. $$

Then we have the following graph.

$$ a_m \xrightarrow{\textbf{Condorcet Winner}} a_{m-1} \xrightarrow{\textbf{Pareto}} a_{m-2} \xrightarrow{\textbf{Pareto}} \cdots \xrightarrow{\textbf{Pareto}} a_1. $$

Similarly, we have

$$ \mathbb{P}[f(P') = a_m] \geqslant \alpha\beta^{m-2} \cdot \mathbb{P}[f(P') = a_1]. $$

Notice that $|\{j \in N : \succ_j \neq \succ'_j\}| = n$. Therefore,

$$
\begin{aligned}
\mathbb{P}[f(P) = a_1] &\geqslant \alpha\beta^{m-2} \cdot \mathbb{P}[f(P) = a_m] \\
&\geqslant \mathrm{e}^{-n\epsilon} \cdot \alpha\beta^{m-2} \cdot \mathbb{P}[f(P') = a_m] &&(\epsilon\text{-DP}) \\
&\geqslant \mathrm{e}^{-n\epsilon} \cdot \alpha^2\beta^{2m-4} \cdot \mathbb{P}[f(P') = a_1] \\
&\geqslant \mathrm{e}^{-2n\epsilon} \cdot \alpha^2\beta^{2m-4} \cdot \mathbb{P}[f(P) = a_1]. &&(\epsilon\text{-DP})
\end{aligned}
$$

Then $\mathrm{e}^{-2n\epsilon}\alpha^2\beta^{2m-4} \leqslant 1$, i.e., $\alpha\beta^{m-2} \leqslant \mathrm{e}^{n\epsilon}$, which completes the proof. □

**Theorem 3.** *If a neutral voting rule $f\colon \mathcal{L}(A)^n \to \mathcal{R}(A)$ satisfies $\epsilon$-DP, $\beta$-Pareto efficiency, and $\alpha$-Condorcet loser criterion, then $\alpha\beta^{m-2} \leqslant \mathrm{e}^{n\epsilon}$.*

*Proof.* Consider the following profile $P$ $(n = 2k+1)$:

- $k+1$ voters: $a_1 \succ a_2 \succ \cdots \succ a_m$;
- $k$ voters: $a_2 \succ \cdots \succ a_m \succ a_1$.

By definition, we have $w_P[a_1, a_i] = 1$, for all $a_i \in A\backslash\{a_1\}$, which indicates that $\mathrm{CL}(P) = a_1$. Also notice that $w_P[a_i, a_j] = n$ for all $i < j$. Thus, $a_i$ Pareto dominates $a_j$ for all $i < j$. The relations among all alternatives are shown in the following graph.

$$a_1 \xrightarrow{\textbf{Pareto}} a_2 \xrightarrow{\textbf{Pareto}} \cdots \xrightarrow{\textbf{Pareto}} a_{m-1} \xrightarrow{\textbf{Condorcet Loser}} a_m.$$

Since $f$ satisfies $\alpha$-Condorcet loser criterion and $\beta$-Pareto efficiency, we have

$$
\begin{aligned}
\mathbb{P}[f(P) = a_1] &\geqslant \beta \cdot \mathbb{P}[f(P) = a_2] \\
&\geqslant \cdots \\
&\geqslant \beta^{m-2} \cdot \mathbb{P}[f(P) = a_{m-1}] \\
&\geqslant \alpha\beta^{m-2} \cdot \mathbb{P}[f(P) = a_m].
\end{aligned}
$$

Now, consider another profile $P'$, where all voters' preferences are exactly the same:

$$a_m \succ a_{m-1} \succ \cdots \succ a_1.$$

Then we have the following graph.

$$a_m \xrightarrow{\textbf{Pareto}} a_{m-1} \xrightarrow{\textbf{Pareto}} \cdots \xrightarrow{\textbf{Pareto}} a_2 \xrightarrow{\textbf{Condorcet Loser}} a_1.$$

Similarly, we have

$$\mathbb{P}[f(P') = a_m] \geqslant \alpha\beta^{m-2} \cdot \mathbb{P}[f(P') = a_1].$$

Notice that $|\{j \in N : \succ_j \neq \succ'_j\}| = n$. Therefore,

$$
\begin{aligned}
\mathbb{P}[f(P) = a_1] &\geqslant \alpha\beta^{m-2} \cdot \mathbb{P}[f(P) = a_m] \\
&\geqslant \mathrm{e}^{-n\epsilon} \cdot \alpha\beta^{m-2} \cdot \mathbb{P}[f(P') = a_m] &&(\epsilon\text{-DP}) \\
&\geqslant \mathrm{e}^{-n\epsilon} \cdot \alpha^2\beta^{2m-4} \cdot \mathbb{P}[f(P') = a_1] \\
&\geqslant \mathrm{e}^{-2n\epsilon} \cdot \alpha^2\beta^{2m-4} \cdot \mathbb{P}[f(P) = a_1]. &&(\epsilon\text{-DP})
\end{aligned}
$$

Then $\mathrm{e}^{-2n\epsilon}\alpha^2\beta^{2m-4} \leqslant 1$, i.e., $\alpha\beta^{m-2} \leqslant \mathrm{e}^{n\epsilon}$, which completes the proof. □

**Proposition 13.** *Condorcet method satisfies SD-efficiency on $\mathcal{D}_C$.*

*Proof.* Let $P$ be an arbitrarily chosen profile in $\mathcal{D}_C$. Then we only need to proof that there does not exist $\xi \in \mathcal{R}(A)$ that SD-dominates $\mathsf{CM}(P)$.

In fact, if there exists such a $\xi$, we can obtain by definition that

$$\sum_{b \succ_j a} \mathbb{P}[\xi = b] \geqslant \sum_{b \succ_j a} \mathbb{P}[\mathsf{CM}(P) = b], \quad \text{for all } j \in N \text{ and } a \in A.$$

Since for any $a \in A$ that $\mathrm{CW}(P) \succ_j a$, we have

$$\sum_{b \succ_j a} \mathbb{P}[\mathsf{CM}(P) = b] = \mathbb{P}[\mathsf{CM}(P) = \mathrm{CW}(P)] = 1,$$

which indicates that

$$\sum_{b \succ_j a} \mathbb{P}[\xi = b] \geqslant 1.$$

Therefore, for any $a \in A$ that $\mathrm{CW}(P) \succ_j a$, $\mathbb{P}[\xi = a] = 0$. However, according to the definition of $\mathrm{CW}(P)$, each $a \in A$ must be ranked behind $\mathrm{CW}(P)$ in some $\succ_j$. Hence we have $\xi = \mathsf{CM}(P)$, a contradiction. □

**Theorem 4.** *There is no neutral voting rule $f\colon \mathcal{L}(A)^n \to \mathcal{R}(A)$ satisfying $\epsilon$-DP, $\alpha$-Condorcet criterion, and $\gamma$-SD efficiency with $\gamma > \frac{\alpha+m-1-\alpha e^{-n\epsilon}}{\alpha+m-1}$.*

*Proof.* Consider the profile $P$, where all voters' vote are exactly the same, i.e.,

$$a_1 \succ_j a_2 \succ_j \cdots \succ_j a_m, \quad \text{for all } j \in N.$$

It is not hard to see that $\mathrm{CW}(P) = a_1$. Since $f$ satisfies $\alpha$-Condorcet criterion, we have $\mathbb{P}[f(P) = a] \leqslant \mathbb{P}[f(P) = a_1]/\alpha$, for all $a \in A \backslash \{a_1\}$. Therefore,

$$1 = \mathbb{P}[f(P) = a_1] + \sum_{a \in A \backslash \{a_1\}} p[f(P) = a] \leqslant \left(1 + \frac{m-1}{\alpha}\right)\mathbb{P}[f(P) = a_1],$$

i.e., $\mathbb{P}[f(P) = a_1] \geqslant \frac{\alpha}{\alpha+m-1}$. Further, by Equation (1), we have

$$\mathbb{P}[f(P) = a_m] \geqslant e^{-n\epsilon} \cdot \mathbb{P}[f(P) = a_1] \geqslant \frac{\alpha e^{-n\epsilon}}{\alpha+m-1}.$$

However, for profile $P$, the unique SD-efficient lottery is $\mathbb{1}_{a_1}$. In other words, all lotteries $\xi \in \mathcal{R}(A)$ that $\xi \neq \mathbb{1}_{a_1}$ are $\gamma$-SD-dominated by $\mathbb{1}_{a_1}$ with $\gamma > 1$. Further,

$$\frac{\sum_{x \succ y} \mathbb{P}[\mathbb{1}_{a_1} = x]}{\sum_{x \succ y} \mathbb{P}[f(P) = x]} \geqslant \frac{\inf_{y \in A} \sum_{x \succ y} \mathbb{P}[\mathbb{1}_{a_1} = x]}{\sup_{y \in A} \sum_{x \succ y} \mathbb{P}[f(P) = x]}$$

$$= \frac{1}{1 - \mathbb{P}[f(P) = a_m]}$$

$$\geqslant \frac{1}{1 - \frac{\alpha e^{-n\epsilon}}{\alpha+m-1}}$$

$$= \frac{\alpha+m-1-\alpha e^{-n\epsilon}}{\alpha+m-1},$$

i.e., $\mathbb{1}_{a_1}$ can $\frac{\alpha+m-1-\alpha e^{-n\epsilon}}{\alpha+m-1}$-dominates $f(P)$, which completes the proof. □

**Theorem 5.** *There is no neutral voting rule $f\colon \mathcal{L}(A)^n \to \mathcal{R}(A)$ satisfying $\epsilon$-DP, $\eta$-Condorcet loser criterion, and $\gamma$-SD-efficiency with $\gamma > \frac{\mathrm{e}^{n\epsilon} - \eta}{\mathrm{e}^{n\epsilon}}$.*

*Proof.* Let $m > 2$. Consider the following profile $P$ with $k(m-2)$ voters ($k \geqslant 1$).

- $k$ voters: $y \succ a_1 \succ a_2 \succ \cdots \succ x \succ a_{m-2}$,
- $k$ voters: $y \succ a_2 \succ a_3 \succ \cdots \succ x \succ a_1$,
- $k$ voters: $y \succ a_3 \succ a_4 \succ \cdots \succ x \succ a_2$,
- $k$ voters: $\cdots$,
- $k$ voters: $y \succ a_{m-2} \succ a_1 \succ \cdots \succ x \succ a_{m-3}$.

Then it is quite evident that $\mathrm{CL}(P) = x$. Since $f$ satisfies $\eta$-Condorcet loser criterion, we have $\mathbb{P}[f(P) = a] \geqslant \eta \cdot \mathbb{P}[f(P) = x]$, for all $a \in A \setminus \{x\}$. By Equation (1), we have $\mathbb{P}[f(P) = x] \geqslant \mathrm{e}^{-n\epsilon}$. Since $f$ satisfies $\eta$-Condorcet loser criterion,

$$\mathbb{P}[f(P) = a] \geqslant \eta \cdot \mathbb{P}[f(P) = x] \geqslant \alpha \mathrm{e}^{-n\epsilon}, \quad \text{for all } a \in A \setminus \{x\}.$$

However, the unique SD-efficient lottery of $P$ is $\mathbb{1}_y$, since $\mathbb{1}_y$ can SD-dominates any other lotteries on $A$. Further,

$$
\begin{aligned}
\frac{\sum\limits_{b \succ a} \mathbb{P}[\mathbb{1}_y = b]}{\sum\limits_{b \succ a} \mathbb{P}[f(P) = b]} &\geqslant \frac{\inf\limits_{a \in A} \sum\limits_{b \succ a} \mathbb{P}[\mathbb{1}_y = b]}{\sup\limits_{a \in A} \sum\limits_{b \succ a} \mathbb{P}[f(P) = b]} \\
&= \frac{1}{1 - \inf\limits_{1 \leqslant i \leqslant m-2} \mathbb{P}[f(P) = a_i]} \\
&\geqslant \frac{1}{1 - \alpha \mathrm{e}^{-n\epsilon}}.
\end{aligned}
$$

In other words, $\mathbb{1}_y$ can $\frac{\mathrm{e}^{n\epsilon} - \eta}{\mathrm{e}^{n\epsilon}}$-SD-dominates $f(P)$, which completes the proof. $\qquad\square$

**Theorem 6.** *There is no neutral voting rule $f\colon \mathcal{L}(A)^n \to \mathcal{R}(A)$ satisfying $\epsilon$-DP, $\gamma$-SD-efficiency, and $\beta$-Pareto efficiency with $\gamma > \frac{\mathrm{e}^{n\epsilon} - \mathrm{e}^{n\epsilon}\beta^{2-m}}{\mathrm{e}^{n\epsilon} - \mathrm{e}^{n\epsilon}\beta^{2-m} + \beta - 1}$.*

*Proof.* Consider the profile $P$, where all voters' preferences are the same, i.e.,

$$a_1 \succ_j a_2 \succ_j \cdots \succ_j a_m, \quad \text{for all } j \in N.$$

By definition, for all $i < j$, any $a_i$ Pareto dominates $a_j$ in profile $P$. In other words, we have the following diagram

$$a_1 \xrightarrow{\mathbf{Pareto}} a_2 \xrightarrow{\mathbf{Pareto}} \cdots \xrightarrow{\mathbf{Pareto}} a_m.$$

Since $f$ satisfies $\beta$-Pareto efficiency, $\mathbb{P}[f(P) = a_{i+1}] \leqslant \beta \cdot \mathbb{P}[f(P) = a_i]$ holds for any $i < m$. By Equation (1), $\mathbb{P}[f(P) = a_m] \geqslant \mathrm{e}^{-n\epsilon} \cdot \mathbb{P}[f(P) = a_1]$. Further,

$$\mathbb{P}[f(P) = a_1] \leqslant \mathrm{e}^{n\epsilon} \cdot \mathbb{P}[f(P) = a_m],$$

$$\mathbb{P}[f(P) = a_2] \leqslant \frac{1}{\beta} \cdot \mathbb{P}[f(P) = a_1] \leqslant \frac{\mathrm{e}^{n\epsilon}}{\beta} \cdot \mathbb{P}[f(P) = a_m],$$

$$\cdots$$

$$\mathbb{P}[f(P) = a_{m-1}] \leqslant \frac{1}{\beta^{m-2}} \cdot \mathbb{P}[f(P) = a_1] \leqslant \frac{\mathrm{e}^{n\epsilon}}{\beta^{m-2}} \cdot \mathbb{P}[f(P) = a_m].$$

By summing up the above inequalities, we have

$$1 = \sum_{a \in A} \mathbb{P}[f(P) = a] \leqslant \left(1 + \left(1 + \frac{1}{\beta} + \cdots + \frac{1}{\beta^{m-2}}\right) \mathrm{e}^{n\epsilon}\right) \cdot \mathbb{P}[f(P) = a_m],$$

i.e., $\mathbb{P}[f(P) = a_m] \geqslant \frac{\beta-1}{\mathrm{e}^{n\epsilon} - \mathrm{e}^{n\epsilon}\beta^{2-m} + \beta - 1}$. However, the unique SD-efficient lottery of $P$ is $\mathbb{1}_{a_1}$, since it can SD-dominate any other lottery. Further, we have

$$\frac{\sum_{b \succ a} \mathbb{P}[\mathbb{1}_y = b]}{\sum_{b \succ a} \mathbb{P}[f(P) = b]} \geqslant \frac{\inf_{a \in A} \sum_{b \succ a} \mathbb{P}[\mathbb{1}_y = b]}{\sup_{a \in A} \sum_{b \succ a} \mathbb{P}[f(P) = b]}$$

$$= \frac{1}{1 - \mathbb{P}[f(P) = a_m]}$$

$$\geqslant \frac{1}{1 - \frac{\beta-1}{\mathrm{e}^{n\epsilon} - \mathrm{e}^{n\epsilon}\beta^{2-m} + \beta - 1}}.$$

In other words, $\mathbb{1}_{a_1}$ can $\frac{\mathrm{e}^{n\epsilon} - \mathrm{e}^{n\epsilon}\beta^{2-m}}{\mathrm{e}^{n\epsilon} - \mathrm{e}^{n\epsilon}\beta^{2-m} + \beta - 1}$-SD-dominates $f(P)$, which completes the proof. $\qquad\square$

## C  More Experimental Figures

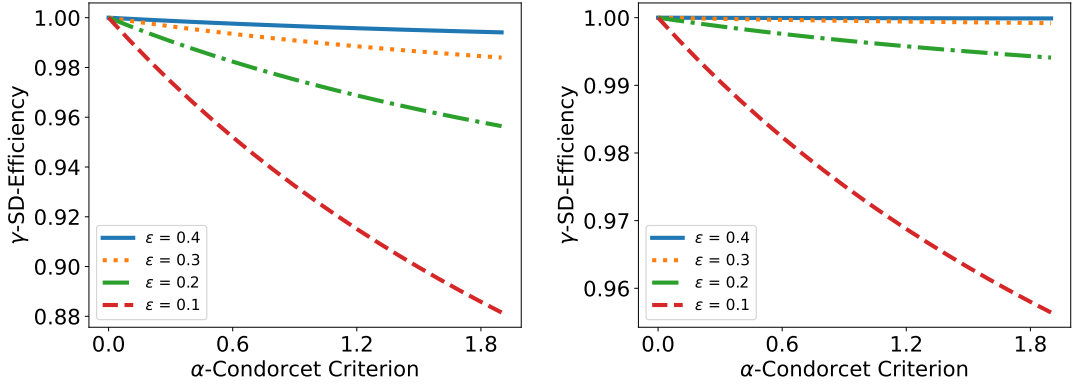

**Fig. 1.** Tradeoff curves between $\alpha$-Condorcet criterion and $\gamma$-SD-efficiency under $\epsilon$-DP (upper bounds). Left: $m = 5, n = 10$. Right: $m = 5, n = 20$.

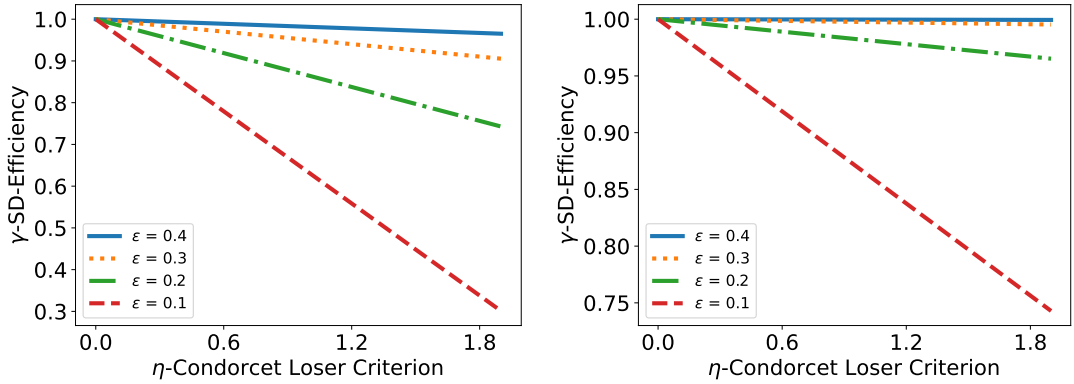

**Fig. 2.** Tradeoff curves between $\eta$-Condorcet loser criterion and $\gamma$-SD-efficiency under $\epsilon$-DP (upper bounds). Left: $m = 5, n = 10$. Right: $m = 5, n = 20$.



\end{document}